\documentclass[superscriptaddress,onecolumn,nofootinbib,prl,amsmath]{revtex4}
\usepackage{times}
\usepackage{graphicx}
\pdfoutput=1
\usepackage{epstopdf}
\usepackage{longtable}
\usepackage{color}
\usepackage{amsmath}
\usepackage{amssymb}
\usepackage[colorlinks,linkcolor=red,citecolor=blue]{hyperref}

\newcommand{\beq}{\begin{equation}}
\newcommand{\eeq}{\end{equation}}
\newcommand{\beqa}{\begin{eqnarray}}
\newcommand{\eeqa}{\end{eqnarray}}
\newcommand{\ket} [1] {\vert#1\rangle}
\newcommand{\bra} [1] {\langle#1\vert}
\newcommand{\braket}[2]{\langle #1 | #2 \rangle}

\def\ket#1{|#1\rangle}
\def\bra#1{\langle #1|}
\def\braket#1#2{\langle #1|#2\rangle}

\def\expect#1{\langle\, #1\, \rangle}

\begin{document}

\title{Observing optical coherence across Fock layers with weak-field homodyne detectors}
\author{Gaia Donati}
\altaffiliation{Corresponding author: gaia.donati@physics.ox.ac.uk}
\author{Tim J. Bartley}
\affiliation{Clarendon Laboratory, University of Oxford, Parks Road, Oxford OX1 3PU, United Kingdom}
\author{Xian-Min Jin}
\affiliation{Clarendon Laboratory, University of Oxford, Parks Road, Oxford OX1 3PU, United Kingdom}
\affiliation{Department of Physics, Shanghai Jiao Tong University, Shanghai 200240, People's Republic of China}
\author{Mihai D. Vidrighin}
\affiliation{Clarendon Laboratory, University of Oxford, Parks Road, Oxford OX1 3PU, United Kingdom}
\affiliation{QOLS, Blackett Laboratory, Imperial College London, London SW7 2BW, United Kingdom}
\author{Animesh Datta}
\affiliation{Clarendon Laboratory, University of Oxford, Parks Road, Oxford OX1 3PU, United Kingdom}
\affiliation{Department of Physics, University of Warwick, Coventry CV4 7AL, United Kingdom}
\author{Marco Barbieri}
\affiliation{Clarendon Laboratory, University of Oxford, Parks Road, Oxford OX1 3PU, United Kingdom}
\affiliation{Dipartimento di Scienze, Universit\`a degli Studi Roma Tre, Via della Vasca Navale 84, 00146 Rome, Italy}
\author{Ian A. Walmsley}
\affiliation{Clarendon Laboratory, University of Oxford, Parks Road, Oxford OX1 3PU, United Kingdom}

\date{\today}


\begin{abstract}

Quantum properties of optical modes are typically assessed by observing their photon statistics or the distribution of their quadratures. Both particle- and wave-like behaviours deliver important information, and each may be used as a resource in quantum-enhanced technologies. Weak-field homodyne detection provides a scheme which combines the wave- and particle-like descriptions. Here we show that it is possible to observe a wave-like property such as the optical coherence across Fock basis states in the detection statistics derived from discrete photon counting. We experimentally demonstrate these correlations using two weak-field homodyne detectors on each mode of two classes of two-mode entangled states. Furthermore, we theoretically describe the response of weak-field homodyne detection on a two-mode squeezed state in the context of generalised Bell inequalities. Our work demonstrates the potential of this technique as a tool for hybrid continuous/discrete-variable protocols on a phenomenon that explicitly combines both approaches.

\end{abstract}



\maketitle


Quantum correlations play a central role in our understanding of fundamental quantum physics and represent a key resource for quantum technologies~\cite{Schroedinger,Wiseman,Modi,Silva}. Progress in quantum information science has followed an increasingly thorough understanding of how such correlations manifest themselves, and how they can be successfully generated, manipulated and characterised~\cite{Baumgratz}. In quantum optical systems, these correlations appear in either properties of the fields, such as quadrature entanglement, or at the level of individual quanta, {\it i.e.}, photons. Access and control over such correlations are key to applications in quantum information science; however, how the complementary mode- and particle-correlations precisely act as a resource is still a subject of debate~\cite{OIreview}. The capability of studying the correlations in a regime traversing mode and particle aspects is thus necessary for clarifying the origin of quantum enhancement. 

When a coherent superposition of many photons occupies a single mode, a wave-like description of the quantum state in terms of continuous variables (\textit{i.e.}, the values of the quadratures) of the electromagnetic field is the standard approach~\cite{Braunstein}. The canonical technique for measuring such light fields is strong-field homodyne detection, which directly probes the quadratures of a field and can provide a full reconstruction of its quantum state~\cite{Lvovsky}. On the other hand, particle-like properties can be directly accessed with a range of photodetectors, a notable example being photon-number-resolving detectors (PNRDs)~\cite{Divochiy,Achilles1,Jaha,Calkins}, and suitable techniques for the reconstruction of the photon statistics~\cite{Zambra,Allevi}. However, such photon counters are intrinsically insensitive to phase, and thus cannot access any coherence between modes. Weak-field homodyne (WFH) has been proposed as a measurement technique bridging wave-like and particle-like descriptions~\cite{Wally,Banaszek,Kuzmich,Resch1,Puentes,Laiho,Zhang}. As in standard homodyne, a local oscillator (LO) provides phase sensitivity, while the photon statistics are accessed by number-resolving detectors. The main difference is that WFH makes use of a classical reference whose mean photon number is of a similar order as that of the probed signal; this allows for the combination of the homodyne technique with PNRDs based on photon counting modules.

In this paper, we employ WFH detection to investigate coherence between different photon-number basis states (Fock layers) across two-mode entangled states. Our detection scheme accesses this manifestation of optical coherence directly, without the need for resource-intensive full state tomography. We demonstrate the oscillations of an array of multi-photon coincidence counts when a split single-photon state (SSPS) and a two-mode squeezed state (TMSS) are interfered with a weak local oscillator. Our experiment can be regarded as a first step towards the quantitative study of the nonlocal properties of multi-mode quantum states with multi-outcome measurements by non-Gaussian detectors. To this end, we theoretically study a violation of Bell inequalities that can be achieved for low numbers of detected photons under ideal conditions. 


\section*{RESULTS}
\begin{figure}[t]
\includegraphics[viewport = -25 220 760 430, clip, width=\columnwidth]{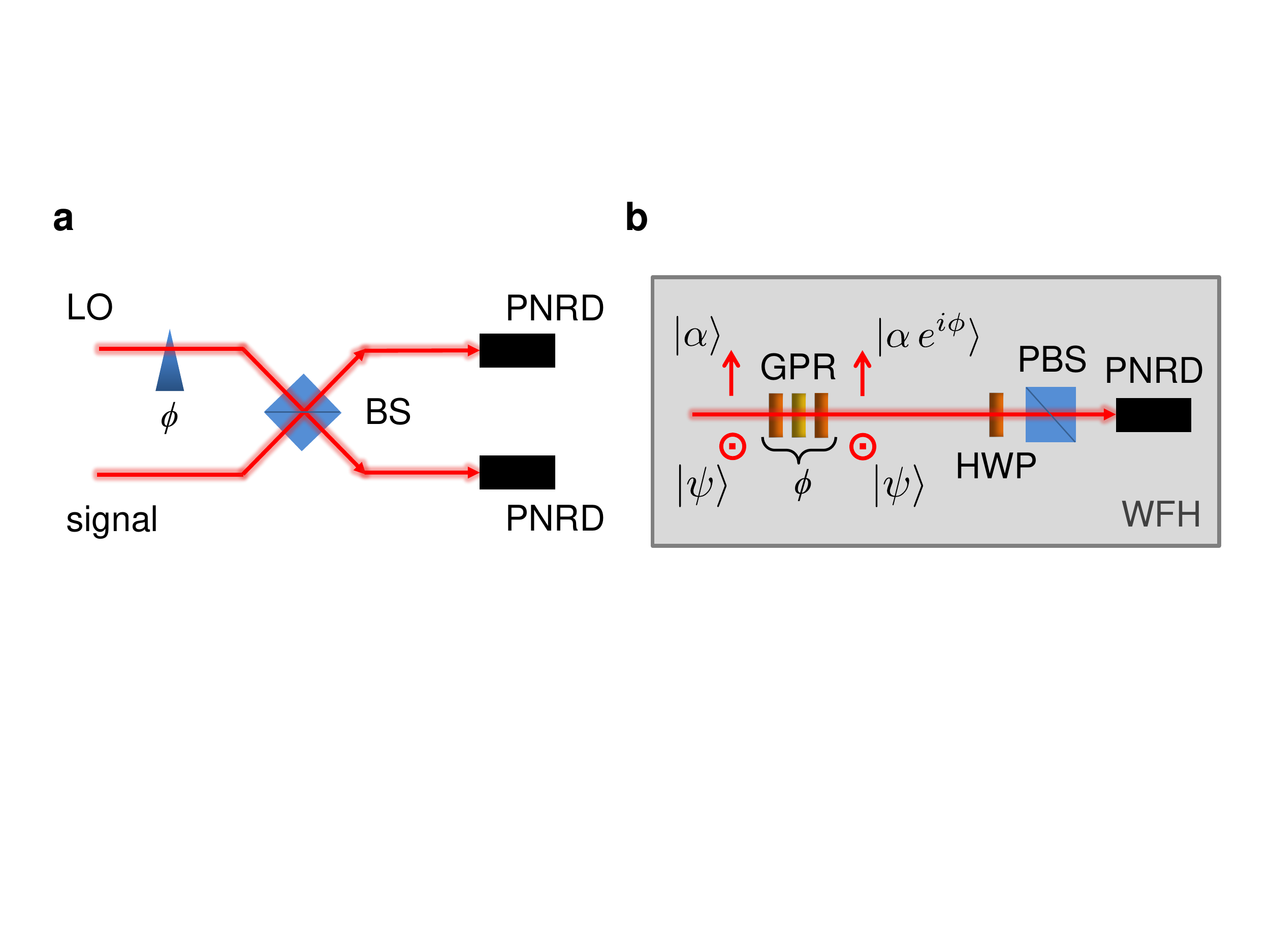}
\caption{{\bf Layout of the weak-field homodyne (WFH) detector.} (a) The general scheme relies on the interference between a signal and a local oscillator (LO) of similar intensity for different settings of the phase $\phi$. Either one or both outputs are measured with photon-number-resolving detectors (PNRDs). (b) Our experimental implementation adopts a collinear configuration in which the signal $\ket{\psi}$ and the LO have orthogonal polarisations. For this reason, the beam splitter (BS) in (a) is replaced by a half-wave plate (HWP) and a polarising beam splitter (PBS) which realise interference in polarisation. The phase setting of the LO is controlled by means of a geometric phase rotator (GPR), which consists of a quarter-wave plate (QWP), a half-wave plate and a second QWP. The rotation of the HWP (with both QWP fixed at $45^\circ$) applies a phase shift to the coherent state. A time-multiplexed detector (TMD) records clicks on the transmitted output mode.}
\label{fig:setup}
\end{figure}

{\bf Experiment}. The weak-field homodyne setup is depicted in Fig.~\ref{fig:setup}a. A signal is mixed on a beam splitter with a local oscillator, described by the coherent state $\ket{\alpha e^{i\phi}}$, which establishes a phase reference. One or both output modes from the beam splitter are then detected using photon-number-resolving detectors. The method is termed ``weak-field'' since the intensity of the LO is comparable to that of the signal. This differs from strong-field homodyne in which the coherent state is many orders of magnitude more intense than the signal, and the outputs are detected by linear photodiodes.

To understand how WFH can detect coherence across Fock layers, consider the detection of $m$ photons. These can originate from any Fock term $\ket{k}$ in the signal together with $m{-}k$ photons from the LO; if the complex coefficients of the terms $\{\ket{k}\}$ have well-defined relative phases, then these result in a modulation of the detection probability $p(m)$ as the phase of the local oscillator is varied. For a two-mode state, the relevant phases are the ones between joint detection events coming from terms in the signal of the form $\ket{k_1}\ket{k_2}$. Consequently, coherence across such events can be observed in the modulation of joint detection probabilities.

We use the layout shown in Fig.~\ref{fig:setup}b to study the coherence between Fock layers across the two modes of a split single photon and a squeezed vacuum state. To maximise the passive phase stability of our setup, we adopt a compact design in which the LO and the signal are collinear and occupy two orthogonal polarisations of a single spatial mode. Our PNRDs are time-multiplexed detectors (TMDs) that split an incoming beam into two spatial and four temporal modes, thereby resolving up to eight photons probabilistically using two avalanche photodiodes~\cite{Achilles1,Lundeen}. Specifically, the TMDs allow us to decompose the intensity of the interference patterns resulting from each output mode into its constituent photon components. In this way we can probe pair-wise correlations between individual Fock layers as we build up a joint detection statistics matrix, every row and column representing the number of clicks in each detected mode. The click statistics gives us access to higher-order Fock states ($k \geq 1$), although this detection scheme is not fully equivalent to a number-resolving detector~\cite{Sperling}.

\begin{figure}[h!]
\includegraphics[viewport = -55 210 790 450, clip, width=\columnwidth]{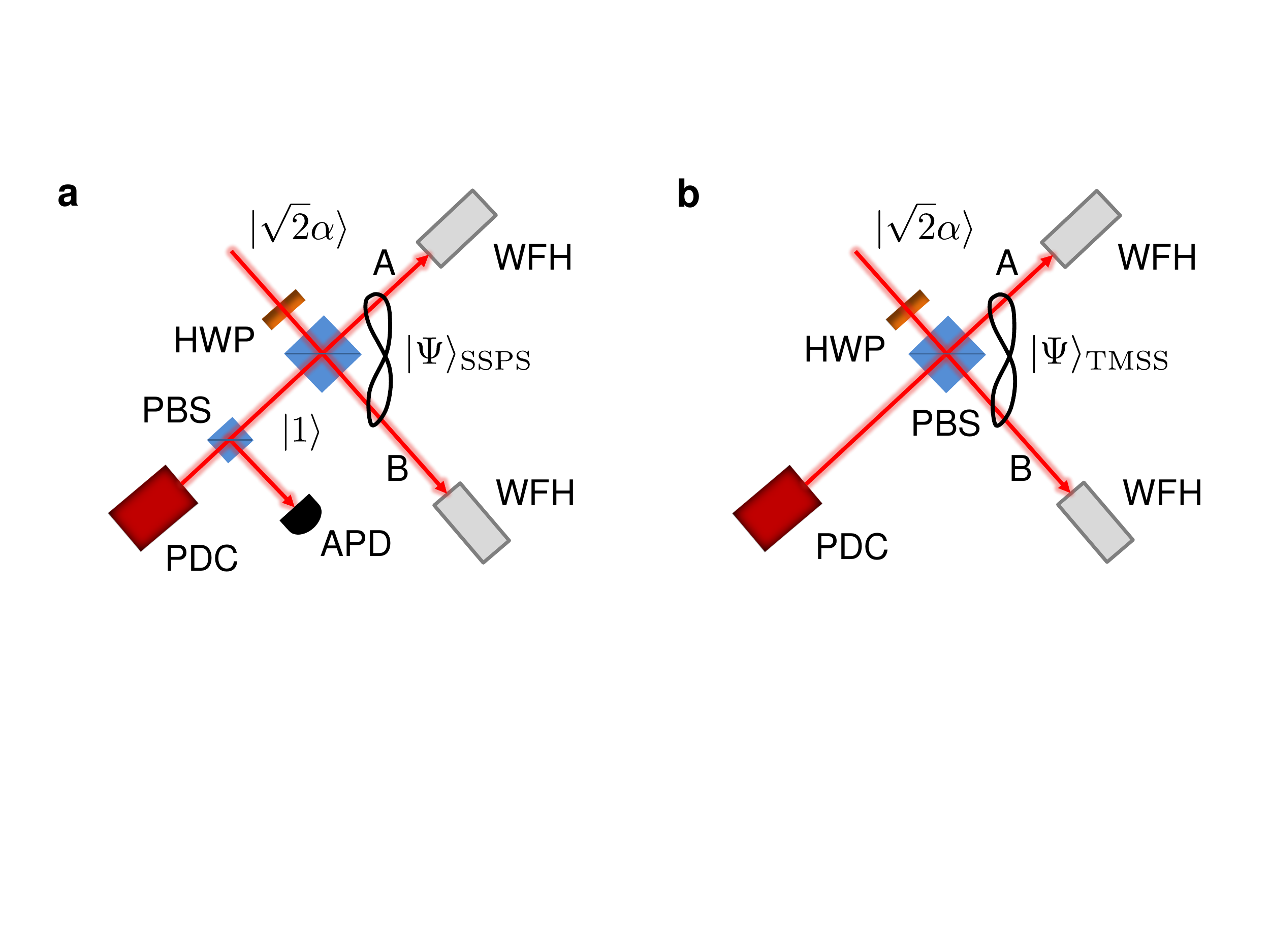}
\caption{{\bf Probing correlations across optical modes with WFH.} (a) A single photon is produced by a heralded source based on parametric downconversion (PDC). We then split the single-photon state in order to generate $\ket{\Psi}_\textrm{SSPS}$ across two separate spatial modes (A and B), both probed by the WFH detectors described in Fig.~\ref{fig:setup}b. (b) The same downconversion source produces the squeezed vacuum state, whose two modes are also probed by weak-field homodyne detectors.}
\label{fig:experiment}
\end{figure}

\begin{figure}[!!!!!h]
\includegraphics[viewport = -100 0 490 290, clip, width=\columnwidth]{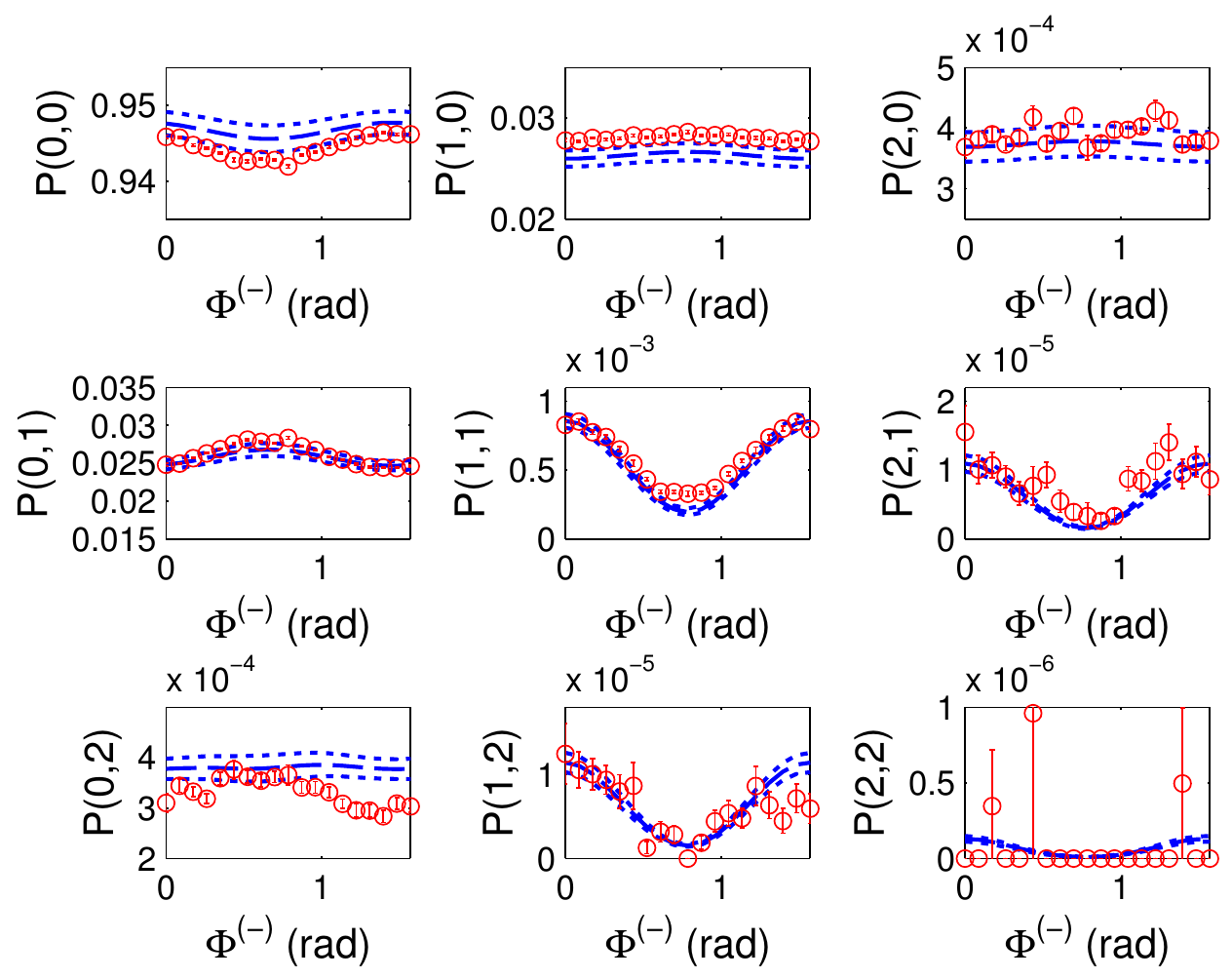}
\caption{{\bf Joint counting statistics for the SSPS.} Correlations between the responses of two WFH detectors as a function of the phase difference $\Phi^{(-)} = \phi^{(-)}/4$ (where $\Phi^{(-)}$ is determined by the settings of the half-wave plates in the phase rotators). We recall that $P(m,m')$ is associated to the joint detection probability of $m$ photons on mode A and $m'$ photons on mode B. The red dots are measured probabilities, with uncertainties determined by a Monte Carlo routine under the assumption of a Poissonian distribution for the raw counts. The blue theoretical curves are obtained from a model which includes the main imperfections in the setup. The dashed curves correspond to an input state with the experimentally determined weights $w_0{=}0.161$ and $w_1{=}0.669$. The uncertainties $\sigma_0{=}0.011$ on $w_0$ and $\sigma_1{=}0.029$ on $w_1$ are estimated by a Monte Carlo routine; the dotted curves show the results given by the theoretical model with $(w_0 - \sigma_0, w_1 - \sigma_1)$ and $(w_0 + \sigma_0, w_1 + \sigma_1)$, respectively. The quantum efficiencies of the detectors, $\eta_{\rm A} = 0.072$ and $\eta_{\rm B} = 0.064$, are experimentally estimated with the Klyshko method~\cite{Klyshko,Achilles2}. The intensities of the two local oscillators are $|\alpha_{\rm A}|{=0.510}$ and $|\alpha_{\rm B}|{=0.585}$.}
\label{fig:SSPS}
\end{figure}

A split single-photon state may be written as
\begin{equation}
\label{Eq:SSPS}
\ket{\Psi}_\textrm{SSPS}=\frac{1}{\sqrt{2}}\bigl(\ket{0}_\textrm{A}\ket{1}_\textrm{B}+\ket{1}_\textrm{A}\ket{0}_\textrm{B}\bigr),
\end{equation}
which describes a single photon in a coherent superposition of modes A and B (see Fig.~\ref{fig:experiment}a and Supplementary Fig.~1 for further details on the experimental layout). The results of our investigation for the SSPS are reported in Fig.~\ref{fig:SSPS}.

The red dots show the experimental probabilities $P(m,m')$ of the joint detection of $m$ clicks in mode A and $m'$ clicks in mode B. Each $P(m,m')$ term is a function of the difference between the phase settings of the weak local oscillators, $\phi^{(-)}{=}\phi_\textrm{A}{-}\phi_\textrm{B}$, since a single photon has no phase {\it per se}~\cite{Lvovsky1,Morin}. Our count rates are such that we only consider events where $m,m'\,{\leq2}$; detection outcomes greater than this level are negligible. The blue lines are predictions from a model which accounts for the following imperfections in our experiment:  non-unit efficiency of the detectors; modulation of the reflectivities of the beam splitters preceding the time-multiplexed detectors when varying the LO phases (due to the geometric phase rotators depicted in Fig.~\ref{fig:setup}b -- see Methods and Supplementary Fig.~2). We also include imperfections in the production of the single-photon states~\cite{Bartley1}, and we thus write 
\begin{equation}
\label{in:state:SSPS}
\rho_0 = w_0\ket{0}\bra{0}+w_1\ket{1}\bra{1}+(1-w_0-w_1)\ket{2}\bra{2},
\end{equation}
where $w_0$ and $w_1$ are coefficients taken between $0$ and $1$ which weight the zero- and one-photon contributions to the input state. We experimentally determine these parameters from the photon statistics of the initial state (see Supplementary Note~1 and Supplementary Note~2). 

The experimental curves show oscillations in the coincidence counts: these are evident in the $P(1,1)$, $P(2,1)$ and $P(1,2)$ terms, as also predicted by the theory. What is most striking about the observed oscillatory behaviour is the fact that it is displayed by terms $P(m,m')$ for which $m + m'\!>\!1$, yet it is determined by the coherence between the terms $\ket{0}_\textrm{A}\ket{1}_\textrm{B}$ and $\ket{1}_\textrm{A}\ket{0}_\textrm{B}$ (see Eq.\eqref{Eq:SSPS}) which do not contain more than one photon each. Ideally, any additional photon detected must therefore come from the local oscillators, hence the SSPS is responsible for the coherent oscillations observed in the considered joint detection probabilities. In practice, we observe good qualitative agreement between the experimental data and our theoretical description. Indeed, we are able to account for all the main features of a 3-by-3 array of multi-photon counts with one model that has no free parameters: detection efficiencies and the weights in Eq.\eqref{in:state:SSPS} are experimentally determined. We attribute the residual discrepancies to two main factors: imperfect mode matching between the LOs and the signal modes, and variations of the laser power during the few-hour long acquisition.

An analogous experiment is conducted on a two-mode squeezed state (see Fig.~\ref{fig:experiment}b and Supplementary Fig.~2 for further details on the experimental layout). This state is archetypal of a quantum resource with well-known correlations between its Fock layers. The expression for a TMSS reads
\begin{equation}\label{Eq:TMSS}
\ket{\Psi}_\textrm{TMSS}=\sqrt{1-|\lambda|^2}\sum_{n=0}^{+\infty}\lambda^n\ket{n}_\textrm{A}\ket{n}_\textrm{B},
\end{equation}
where the real squeezing parameter $|\lambda|$ governs the photon distribution across the photon-number basis states. Here we pump our source with moderate parametric gain in order to generate significant higher-order Fock layers in the two-mode state~\cite{Mosley}. The expected phase dependence of $\{P(m,m')\}$ for a two-mode squeezed state was shown to be $\phi^{(+)}{=}\phi_\textrm{A}{+}\phi_\textrm{B}$~\cite{Blandino}. The TMSS has a phase dependence arising from that of the pump; in a conventional strong homodyne setup this would define which quadratures are squeezed.  In our case, the same effect is manifested in the phase dependence of the click patterns; hence, appropriate phase locking was necessary (see Methods).

\begin{figure}[!!!!!h]
\includegraphics[viewport= -100 0 490 290, clip, width=\columnwidth]{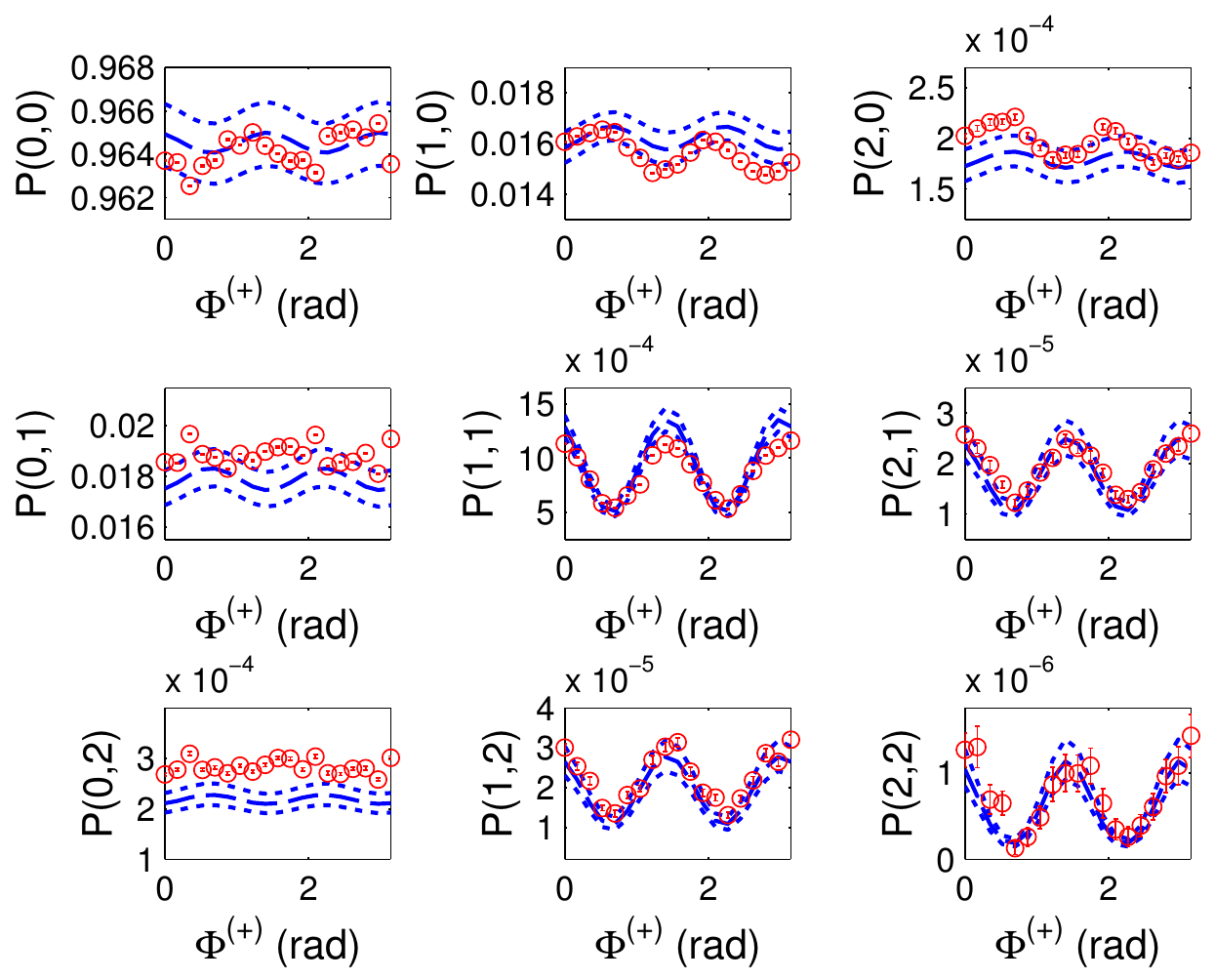}
\caption{{\bf Joint counting statistics for the TMSS.} Correlations between the responses of two WFH detectors as a function of the phase sum $\Phi^{(+)} = \phi^{(+)}/4$ (where $\Phi^{(+)}$ is determined by the settings of the half-wave plates in the phase rotators). We recall that $P(m,m')$ is associated to the joint detection probability of $m$ photons on mode A and $m'$ photons on mode B. The red dots are measured probabilities, with uncertainties determined by a Monte Carlo routine under the assumption of a Poissonian distribution for the raw counts. The blue theoretical curves are obtained from a model which includes the main imperfections in the setup. The dashed curves correspond to an input state with the experimentally determined squeezing parameter $|\lambda|{=}0.295$ and weight $p{=}0.04$ for the additional noise term. The uncertainty $\sigma{=}0.016$ on $|\lambda|$ is estimated by a Monte Carlo routine; the dotted curves show the results  given by the theoretical model with $|\lambda| \pm \sigma$. The experimental quantum efficiencies of the detectors are $\eta_{\rm A} = 0.132$ and $\eta_{\rm B} = 0.155$ (estimated with the Klyshko method). The intensities of the two local oscillators are $|\alpha_{\rm A}|{=0.365}$ and $|\alpha_{\rm B}|{=0.347}$.}
\label{fig:TMSS}
\end{figure}
The results obtained for the two-mode squeezed state are illustrated in Fig.~\ref{fig:TMSS}. Once again, we observe good qualitative agreement between our data and the predictions from a theoretical model that includes the same imperfections as for the single-photon case. Here the input state can be modelled as 
\begin{equation}\label{Eq:rhoTMSS}
\rho = (1 - p)\ket{\Psi}_\textrm{TMSS}\bra{\Psi} + p\ket{0,1}\bra{0,1}\,,
\end{equation}
where the extra term is a first-order approximation of noise in a squeezed thermal state. This asymmetry across the modes is justified by the experimentally recorded $g^{(2)}$ for the parametric downconversion source: we find $g^{(2)}_{\textrm{A}} = 1.98 \pm 0.04$ and $g^{(2)}_{\textrm{B}} = 1.92 \pm 0.04$ for the two marginals, to be compared with $g^{(2)} = 2$ for an ideal thermal state. The lower value of $g^{(2)}_{\textrm{B}}$ motivates the addition of the noise term on output mode B. The quoted values of $g^{(2)}$ also suggest that additional (and undesired) Schmidt modes might be responsible for the presence of photons in modes correlated to undetected modes~\cite{Mosley2}. As for the case of the SSPS, imperfect mode matching between the LOs and the signal modes and variations of the laser power during the data acquisition are recognised as the main causes of the residual departures of the experimental curves from the expected behaviour in Fig.~\ref{fig:TMSS}. \\
As a general remark applying to both classes of studied states, we note that our numerical models depend strongly on the detection efficiencies (see Supplementary Fig.~4 and Supplementary Fig.~5). In this sense, the more pronounced discrepancies in Fig.~\ref{fig:SSPS} and Fig.~\ref{fig:TMSS}, such as that in the $P(0,2)$ term, may be due to noise affecting our estimation of these parameters. More details on the theoretical model for both the SSPS and the TMSS can be found in Supplementary Note~1 and Supplementary Note~2.


\bigskip

{\bf Generalised Bell inequalities for WFH detection.} Correlations such as those revealed in our experiment need to be understood in relation to canonical criteria for non-classicality, for instance the violation of a Bell inequality, to assess their role as possible quantum resources~\cite{Brunner}. The ability to access and discriminate higher photon numbers leads us to refer to generalised, higher-dimensional Bell inequalities~\cite{Collins}, which we study theoretically in the context of our experiment with a TMSS. 

\begin{figure}[h!]
\includegraphics[viewport= -80 180 820 520, clip, width=\columnwidth]{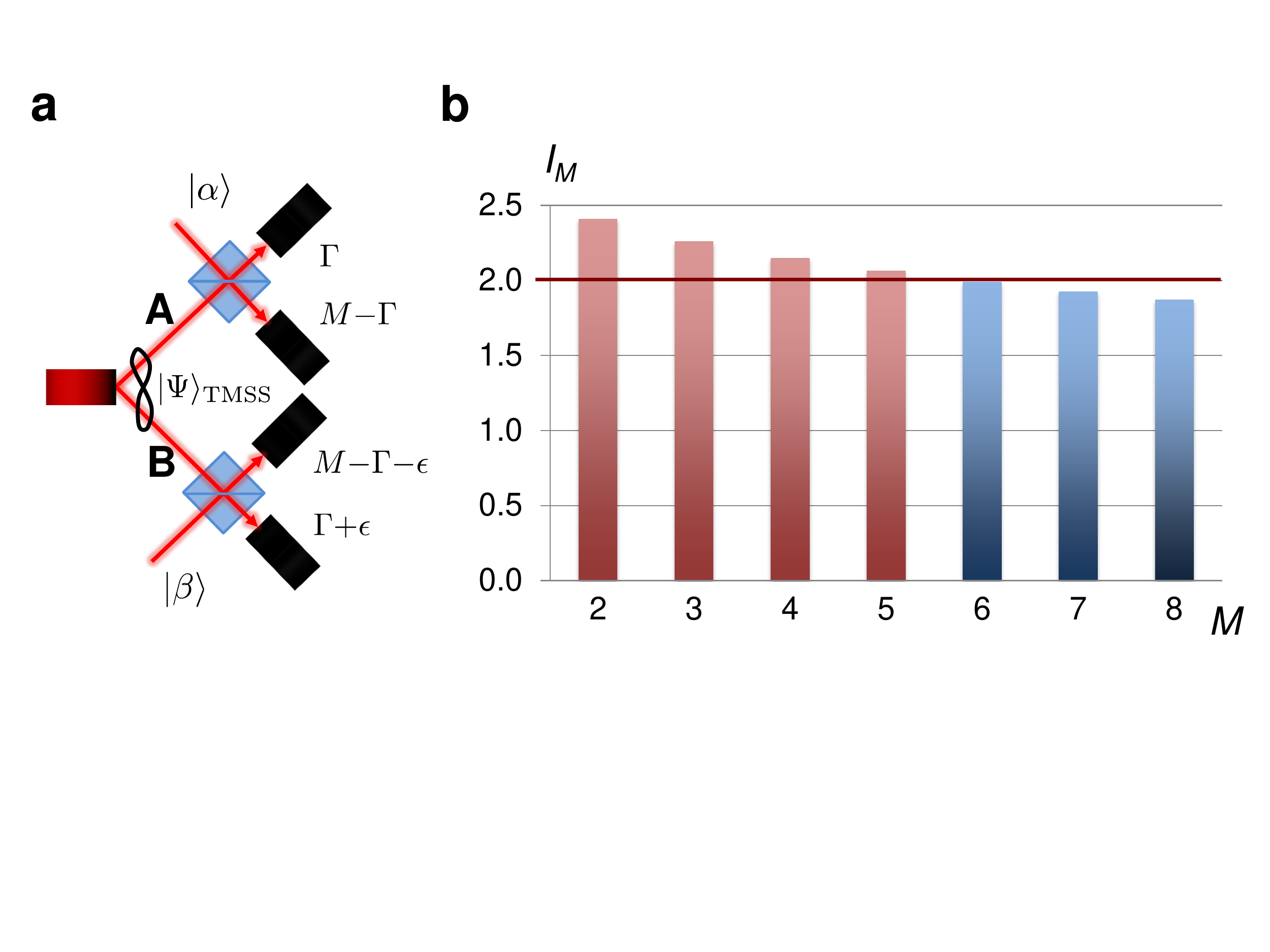}
\caption{{\bf Generalised Bell inequalities with WFH.} (a) In the ideal case, the two modes of a TMSS are interfered with two weak local oscillators; PNRDs monitor all four outputs. We fix the number of detected photons to $M$ on each side, and label the various detection events recorded by the PNRDs according to how the $M$ particles distribute themselves on the two output modes:  $\Gamma, M\!-\!\Gamma, \Gamma\!+\!\epsilon$ and $M\!-\!\Gamma\!-\epsilon.$ (b) The plot shows the values taken by the quantity $I_M$ (see Eq.(\ref{CGLMP})) when the number of detected photons $M$ on each side varies between $2$ and $8$. The bound for $I_M$ predicted by local realism is $2$ for all values of $M$, hence we see that no violation of a generalised Bell inequality is possible beyond $M = 5$.}
\label{fig:theory}
\end{figure}

To this end, let us consider the scenario depicted in Fig.~\ref{fig:theory}. Each mode of a TMSS is analysed by means of a WFH detector; here all four outputs are monitored by perfect PNRDs. We take into account Fock layers which lead to the detection of $M$ photons on each side: the detected photons are split among the two output ports of the PBSs (\textit{i.e.}, the photons are either transmitted or reflected) according to the convention indicated in Fig.~\ref{fig:theory}a. Specifically, we consider measurements with $D{=}M{+}1$ possible results, distinguished according to the number of photons $\Gamma$ detected on the transmitted arm of each PBS. We are thus interested in the probabilities associated with joint detection events comprising outcomes (on each output arm, on both sides) differing by $\epsilon$ (the outcomes being taken modulo $D$). $\{\alpha,\alpha'\}$ and $\{\beta,\beta'\}$ denote the LO settings on side $A$ and side $B$, respectively. The relevant probabilities are combined into the expression
\begin{widetext}
\begin{equation}\label{CGLMP}
\begin{split}
I_M{=}&\sum_{\epsilon{=}0}^{\bigl[\frac{M+1}{2}\bigr]-1}\Biggl(1-\frac{2\epsilon}{M}\Biggr)  \Bigl[P(\alpha,\beta,\epsilon) + P(\alpha',\beta,-\epsilon-1) + P(\alpha',\beta',\epsilon) + P(\alpha,\beta',-\epsilon)\Bigr] \\
& - \Bigl[P(\alpha,\beta,-\epsilon-1) + P(\alpha',\beta,\epsilon) + P(\alpha',\beta',-\epsilon-1) + P(\alpha,\beta',\epsilon+1)\Bigr],
\end{split}
\end{equation}
\end{widetext}
where the local realistic bound is $|I_M|\leq 2$~\cite{Collins}. We note that the case $M{=}1$ corresponds to the standard CHSH inequality which was experimentally tested in~\cite{Kuzmich}. Additional details on how to compute $I_M$ for the specific layout that we consider are provided in Supplementary Note~3, Supplementary Fig.~6 and Supplementary Fig.~7.

We run a numerical search for values of $M$ ranging from $2$ to $8$ in order to find the set of parameters $\{\lambda, \alpha, \alpha',\beta,\beta'\}$ which determine the highest violation of these generalised Bell inequalities. Our results are shown in Fig.~\ref{fig:theory}: no violation of Eq.\eqref{CGLMP} is found beyond $M{=}5$. This behaviour comes from the fact that the maximal violation of Eq.\eqref{CGLMP} (\textit{i.e.}, the attainment of the maximum value allowed by quantum mechanics) relies on a particular structure for the entangled state~\cite{Acin}. Notably, a two-mode squeezed state is specified solely by the squeezing parameter $\lambda$: this means that there are not enough degrees of freedom to tune in order to obtain the required form for the input state. This restriction becomes more severe as the dimensionality of the system increases, to the point that no violation of local realism can be inferred despite the use of an entangled resource such as a TMSS. Hence the decrease in the violation stems from the Gaussian character of the two-mode squeezed state, which fixes the functional shape of the oscillation curves simultaneously for all Fock layers.


\section*{DISCUSSION}
We have shown how WFH can be used to deconstruct phase-dependent measurements on two-mode entangled states into their constituent Fock layers. The ability to operate devices at the interface of wave and particle regimes opens up new possibilities for quantum information processing~\cite{Kiesel}. Thus, our work provides a new insight on such resources within the broader investigations on hybrid continuous/discrete-variable coding~\cite{Alexei,Jonas1,Jonas2,Bellini,Laurat}.

As a complement, we have also studied the theory of the violation of generalised Bell inequalities using an ideal WFH setup. These tests shed light on how the transition from non-Gaussian to Gaussian measurements occurs. Such transition is achieved in the context of WFH detection as an increase in the LO strength. In fact, this study may also be interpreted in the more general framework of Gaussian vs non-Gaussian measurements, where it is well-known that the outcomes of a Gaussian measurement on a Gaussian quantum state can be explained by a local realistic model. For this reason, strong-field homodyne detection on a TMSS cannot be used to violate a Bell inequality~\cite{Grangier,Gilchrist}. On the other hand, WFH is an example of non-Gaussian measurement, as attested by its Wigner function. Consequently, we expect that the outcomes of WFH detection on an entangled Gaussian state cannot be described by a local realistic model~\cite{Kuzmich,Bjoerk,Hessmo}; however, this breaks for moderately high photon numbers. 

On the experimental side, further developments for the observation of the violation of higher-dimensional Bell inequalities demand detectors with higher quantum efficiency~\cite{Eisaman}. These are not only necessary for achieving significant counting statistics, but also to prevent one Fock layer to be affected from higher-order contributions. Encouraging results have been obtained in this direction with cryogenic detectors.


\section*{METHODS}
{\bf Source of quantum states.} A pulsed Ti:Sapphire laser (repetition rate $256$kHz, central wavelength $\lambda_{\rm TiSa} = 830$nm and bandwidth $\Delta\lambda_{\rm TiSa} \simeq 30$nm) is doubled in a nonlinear crystal ($\beta$-barium borate, BBO). The second-harmonic beam ($\lambda_{\rm UV} = 415$nm) pumps a type-II parametric downconversion process in a nonlinear crystal (potassium dihydrogen phosphate, KDP) in order to produce a two-mode vacuum-squeezed state. The source is designed to generate spectrally uncorrelated modes, based on group velocity mismatch inside the birefringent medium~\cite{Mosley}. Daughter photons have orthogonal polarisations and different spectral widths ($\Delta\lambda_{\textrm{V}} \simeq 12$nm, $\Delta\lambda_{\textrm{H}} \simeq 6$nm). The same source generates the split single-photon state and the two-mode squeezed state, depending on the ultra-violet pump beam brightness.

\bigskip

{\bf Weak-field homodyne detection.}  The detection system that we adopt in our experiment is realised with a collinear geometry in order to ensure passive phase stability. Each mode in Eq.  \eqref{Eq:SSPS} and \eqref{Eq:TMSS}  is superposed with a local oscillator of orthogonal polarisation at a polarising beam splitter (PBS). This delivers a common spatial mode at the output but orthogonal polarisations for the signal to be probed and the weak coherent beam. Interference is then realised by a half-wave plate and an additional PBS. The relative phase between the horizontal and vertical polarisations can be adjusted by means of a geometric phase rotator (GPR -- see Fig.~\ref{fig:setup}), which is composed of a quarter-wave plate (QWP), a half-wave plate and a second QWP. The axis of the first QWP is set to $45^\circ$ in order to transform the input linear polarisations into circular ones. The rotation of the HWP by an angle $\phi/4$ results in a phase shift equal to $\phi$ between left- and right-circular polarisations. The initial linear polarisations are then recovered by setting the second QWP to $45^\circ$. The successful calibration of the full device relies on the characterisation of each element (including the PBS), particularly of the QWPs. When the GPR is correctly calibrated power fluctuations around $3\%$ are recorded, while imperfections in the calibration of one of its constituents can lead to fluctuations in power above $10\%$ (see Supplementary Fig.~2). Finally, the output state is analysed with a time-multiplexed detector: this consists of two fibre-based cascaded Mach-Zehnder interferometers that split the incoming state over two spatial modes and four distinct time bins. Time-resolved clicks from avalanche photodiodes monitoring the two transmitted modes are thus registered.

\bigskip

{\bf Active phase stabilisation.} To actively lock the phase set by the GPRs, we use an ancillary laser beam (from a continuous-wave HeNe laser, $\lambda' = 633$nm) which back-propagates through the interferometer. The classical interference pattern that we obtain when the ancillary beam reproduces correctly the signal and LO optical paths constitutes the signal recorded by a photodiode connected to a PID device (SRS SIM$960$ Analog PID Controller).


\section*{ACKNOWLEDGEMENTS}
The authors are grateful to J. Spring and A. Eckstein for helping with the detection setup, and thank L. Zhang, B. Smith, M.S. Kim and M. Paternostro for fruitful discussions. This work was supported by the EPSRC (EP/K034480/1,
EP/H03031X/1, EP/H000178/1), the EC project SIQS and the Royal Society. X.-M. J. acknowledges support from the National Natural Science Foundation of China (Grant No. 11374211) and an EC Marie Curie fellowship (PIIF-GA-2011-300820). A. D. is supported by the EPSRC (EP/K04057X/1).


\section*{AUTHOR CONTRIBUTIONS}
G.D. performed the experiments and, with assistance from A.D., M.B. and M.D.V., analysed the data and investigated the application of WFH to generalised Bell inequalities. G.D. and T.J.B., with assistance from M.B., X.-M.J. and M.D.V., designed the experimental layout. T.J.B., M.B. and I.A.W. conceived the experiment. G.D., T.J.B., A.D. and M.B. wrote the paper with contributions from all authors. All authors agree to the contents of the article.

\newpage

\section*{SUPPLEMENTARY INFORMATION}

\begin{figure}[h!]
\includegraphics[viewport= -80 -10 780 500, clip, width=\columnwidth]{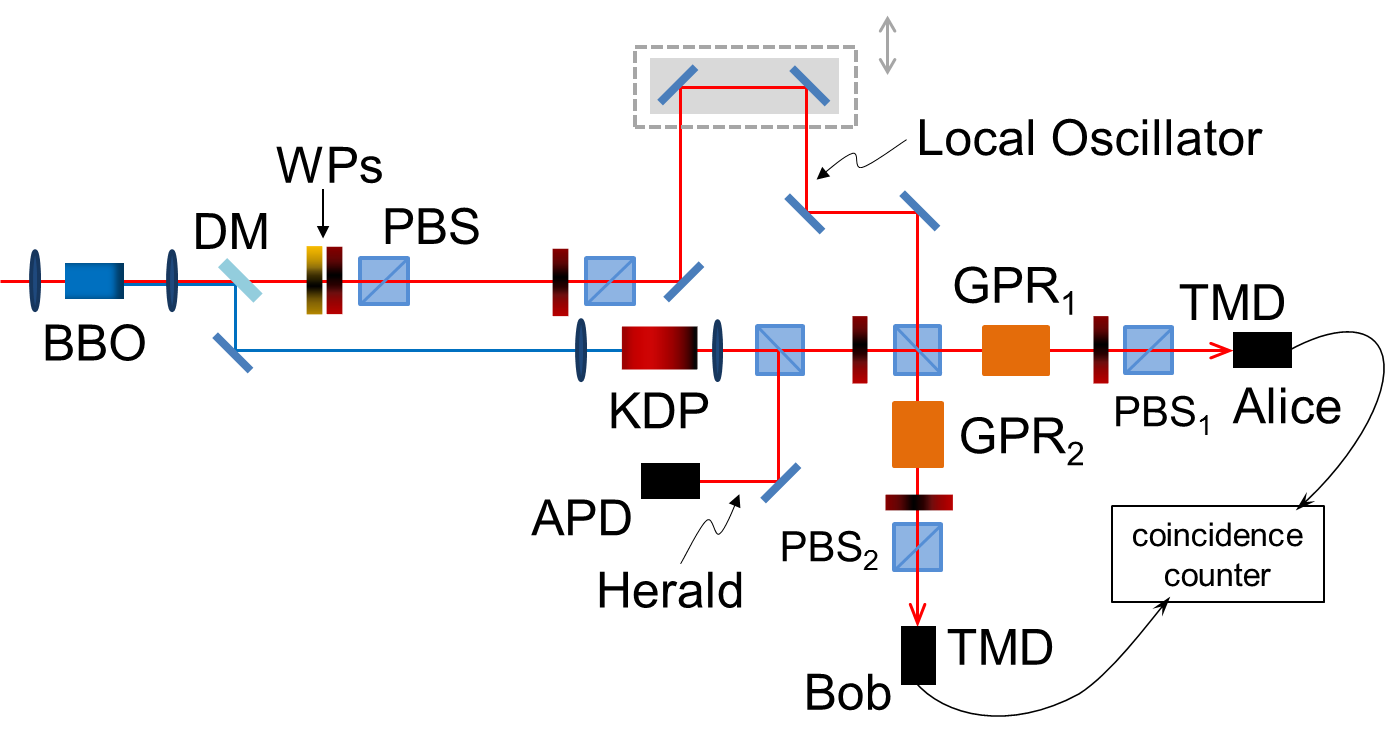}
\flushleft{{\bf Supplementary Figure 1: Experimental layout for the study of optical coherence via weak-field homodyne across the modes of a split single-photon state (SSPS).} A blue beam ($\lambda_{\rm UV} = 415$nm) interacts with a Type-II potassium dihydrogen phosphate birefringent crystal to produce frequency-degenerate downconverted photons with orthogonal polarisations at $830$nm. The ultraviolet beam is obtained via second-harmonic generation on a Type-I $\beta$-barium borate crystal pumped by the radiation coming from a pulsed femtosecond Ti:Sapphire laser system (Coherent Mira Seed -- RegA, central wavelength $\lambda_{\rm TiSa} = 830$nm, repetition rate $256$kHz). The vertically polarised downconversion mode is detected by an avalanche photodiode and thus heralds the presence of a signal on the twin mode, which is mixed with a local oscillator on a polarising beam splitter (PBS). The signal and coherent state, now collinear, are then interfered in polarisation on two PBSs preceded by geometric phase rotators which allow us to vary the phase of the local oscillators independently on Alice's and Bob's side. The two transmitted output modes are monitored by two time-multiplexed detectors, while the reflected ones are discarded. A field-programmable gate array processes the recorded coincidence and single counts. BBO: $\beta$-barium borate crystal -- DM: dichroic mirror -- WP: wave plate (yellow, quarter-wave plate; red, half-wave plate) -- PBS: polarising beam splitter -- KDP: potassium dihydrogen phosphate crystal -- APD: avalanche photodiode -- GPR: geometric phase rotator -- TMD: time-multiplexed detector.}
\end{figure}

\newpage

\begin{figure}[h!]
\includegraphics[viewport= -15 60 730 650, clip, width=\columnwidth]{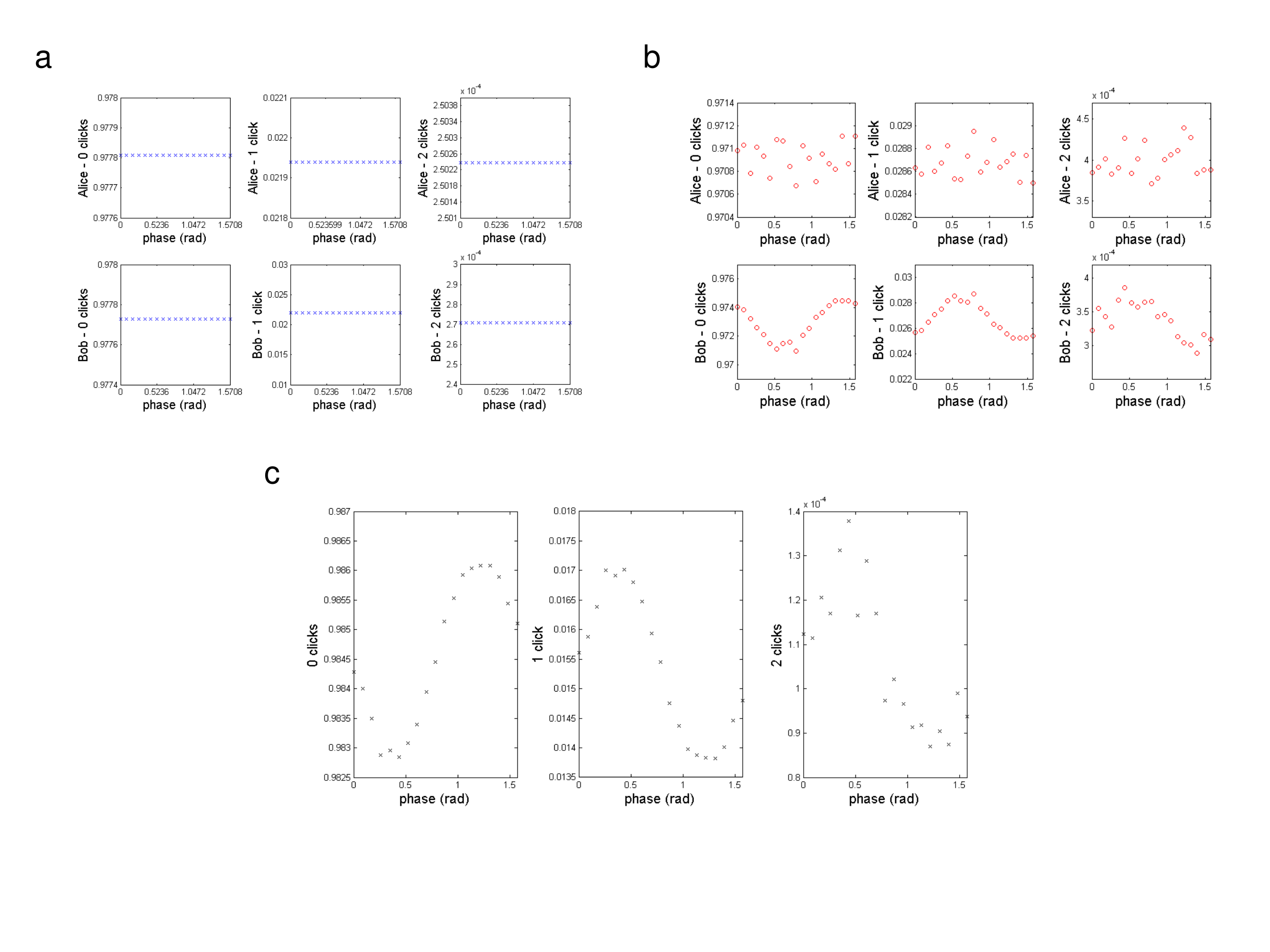}
\flushleft{{\bf Supplementary Figure 2: Effect of an imperfect geometric phase rotator (GPR) on the marginal distributions for detection events up to two photons.} (a) Theoretical marginal distributions for the split single-photon state. (b) Experimental marginal distributions for the SSPS when the GPR on Bob's side is moved. (c) Typical calibration curve for a geometric phase rotator. Theoretically, no term in the marginal counting statistics should oscillate when varying the phase of the local oscillator (see plot (a)). However, the experimental data (see graph (b)) reveals the presence of oscillations in the recorded counts on the mode where the geometric phase rotator is active. Analogous results are found for the TMSS. These plots suggest that a non-ideal GPR can cause spurious oscillations in the marginals, that is, the distributions of detection events when we trace over one mode of the full two-mode output state. The calibration curve (see plot (c)) is obtained by interfering the vacuum with a local oscillator and recording counts on one detector as the GPR is moved. We note that the visibility of the oscillations in the LO counts is about $10\%$ for the $1$-click term. If we compare this value with the visibilities of the oscillations in plot (b), we see that it falls between the $1$-click term ($V \simeq 6\%$) and the $2$-click term ($V > 10\%$ - although in this case the absolute values are of the order of $10^{-4}$ counts $\cdot$ s$^{-1}$) associated to the experimental marginal distributions. These observations support the addition of a modulation of the reflectivity in our theoretical models for the split single photon and the two-mode squeezed state.}
\end{figure}

\newpage

\begin{figure}[h!]
\includegraphics[viewport= -100 -20 780 600, clip, width=\columnwidth]{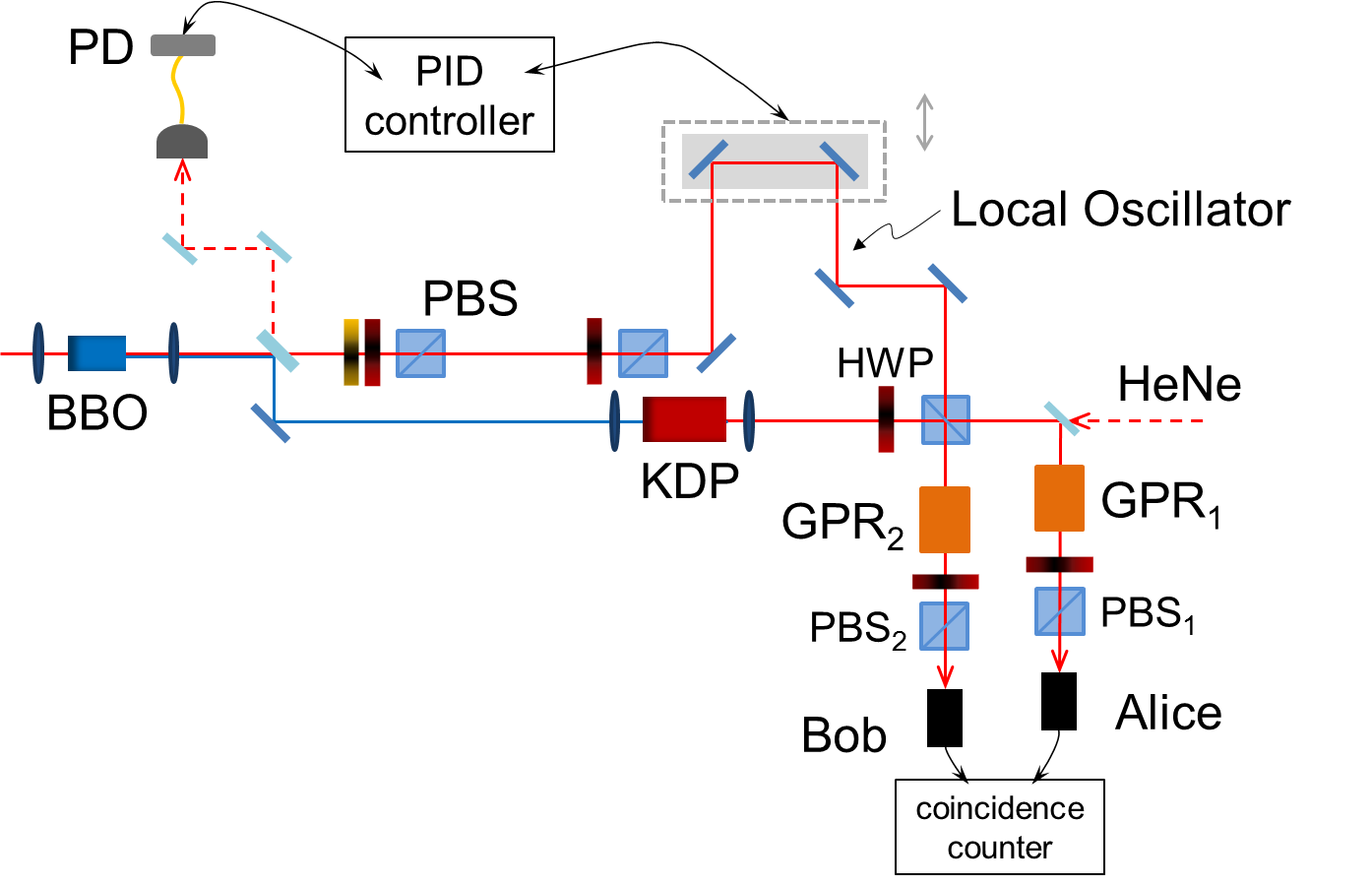}
\flushleft{{\bf Supplementary Figure 3: Experimental layout for the study of optical coherence via weak-field homodyne across the modes of a two-mode squeezed state (TMSS).} The setup for the generation of two-mode squeezed vacuum states exhibits a few modifications with respect to that illustrated in Supplementary Fig~1. In this case we interfere both downconversion modes with a local oscillator (LO); indistinguishability in polarisation is achieved thanks to a half-wave plate and a polarising beam splitter on each output mode. Once again, the geometric phase rotators allow us to vary the phases of the local oscillators. This layout requires active phase stabilisation to ensure that the phase difference between each initial LO phase and the phase of the two-mode squeezed state is kept fixed. Our active phase locking relies on an ancillary laser beam from a continuous-wave HeNe laser ($\lambda' = 633$nm) which back-propagates through the interferometer. The signal produced by the classical interference pattern -- which is obtained when the ancillary beam reproduces correctly both optical paths in the interferometer -- is recorded by a photodiode connected to a PID device (SRS SIM$960$ Analog PID Controller) whose electronic signal is then fed to a piezo delay stage in the LO path (indicated by the gray dashed box in the diagram). BBO: $\beta$-barium borate crystal -- PBS: polarising beam splitter -- HWP: half-wave plate -- KDP: potassium dihydrogen phosphate crystal -- GPR: geometric phase rotator -- HeNe: helium-neon diode laser -- PD: photodiode -- PID: Proportional-Integral-Derivative device.}
\end{figure}

\newpage

\begin{figure}[h!]
\includegraphics[viewport= -150 -10 550 320, clip, width=\columnwidth]{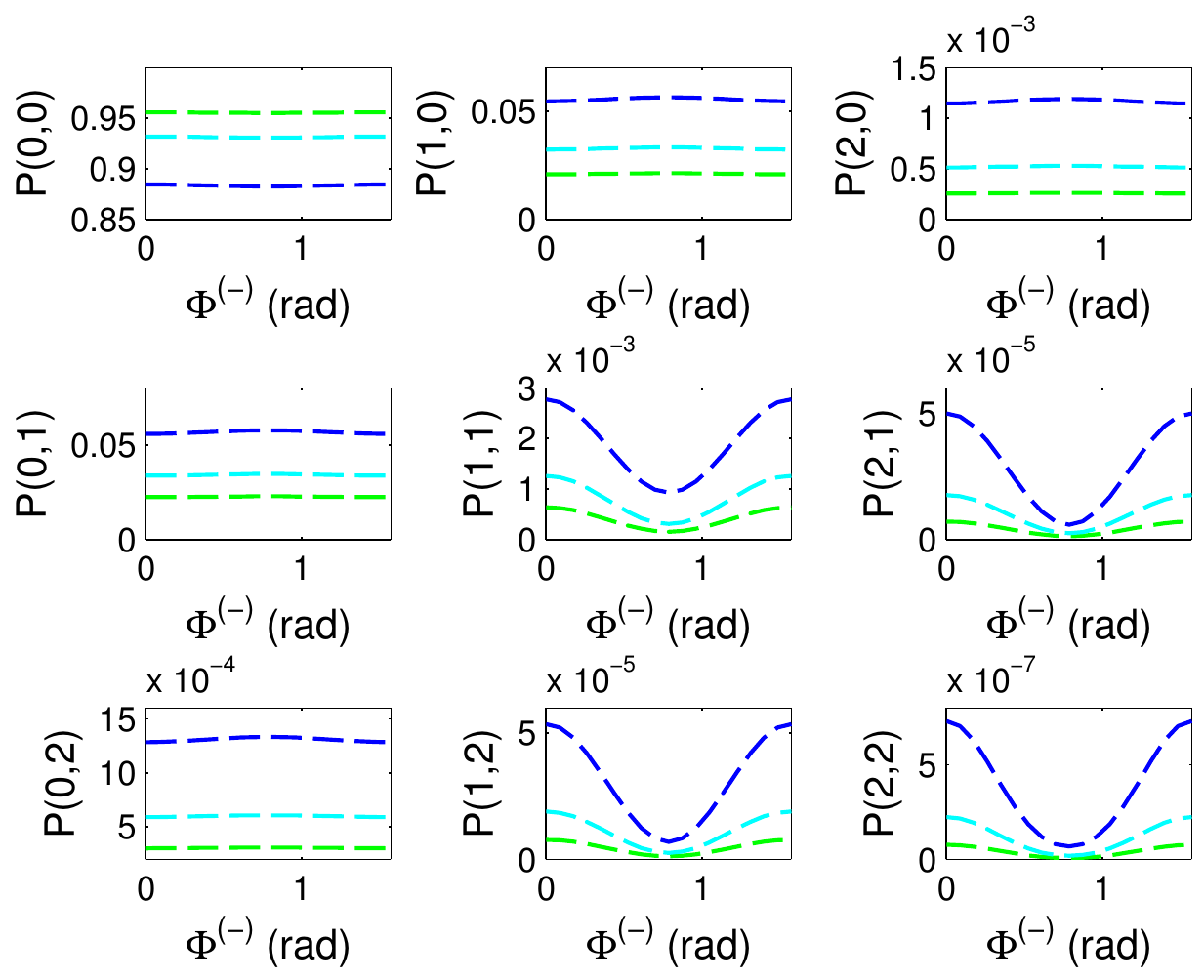}
\flushleft{{\bf Supplementary Figure 4: Theoretical joint counting statistics for different detection efficiencies in the case of a split single-photon state.} To illustrate the effect of imperfect detection on the observed nonclassical oscillations, we plot the array of multi-photon coincidence counts while varying the values of the detection efficiencies. Green curves correspond to $\eta = 0.05$, cyan to $\eta = 0.1$ and blue to $\eta = 0.2$; we assume the same efficiencies on both time-multiplexed detectors and use the experimentally determined values of the coefficients $w_0$ and $w_1$. We see how the parameter $\eta$ changes both the level of detected signal and, most importantly, the visibility of the oscillations.}
\end{figure}

\begin{figure}[h!]
\includegraphics[viewport= -150 -10 540 300, clip, width=\columnwidth]{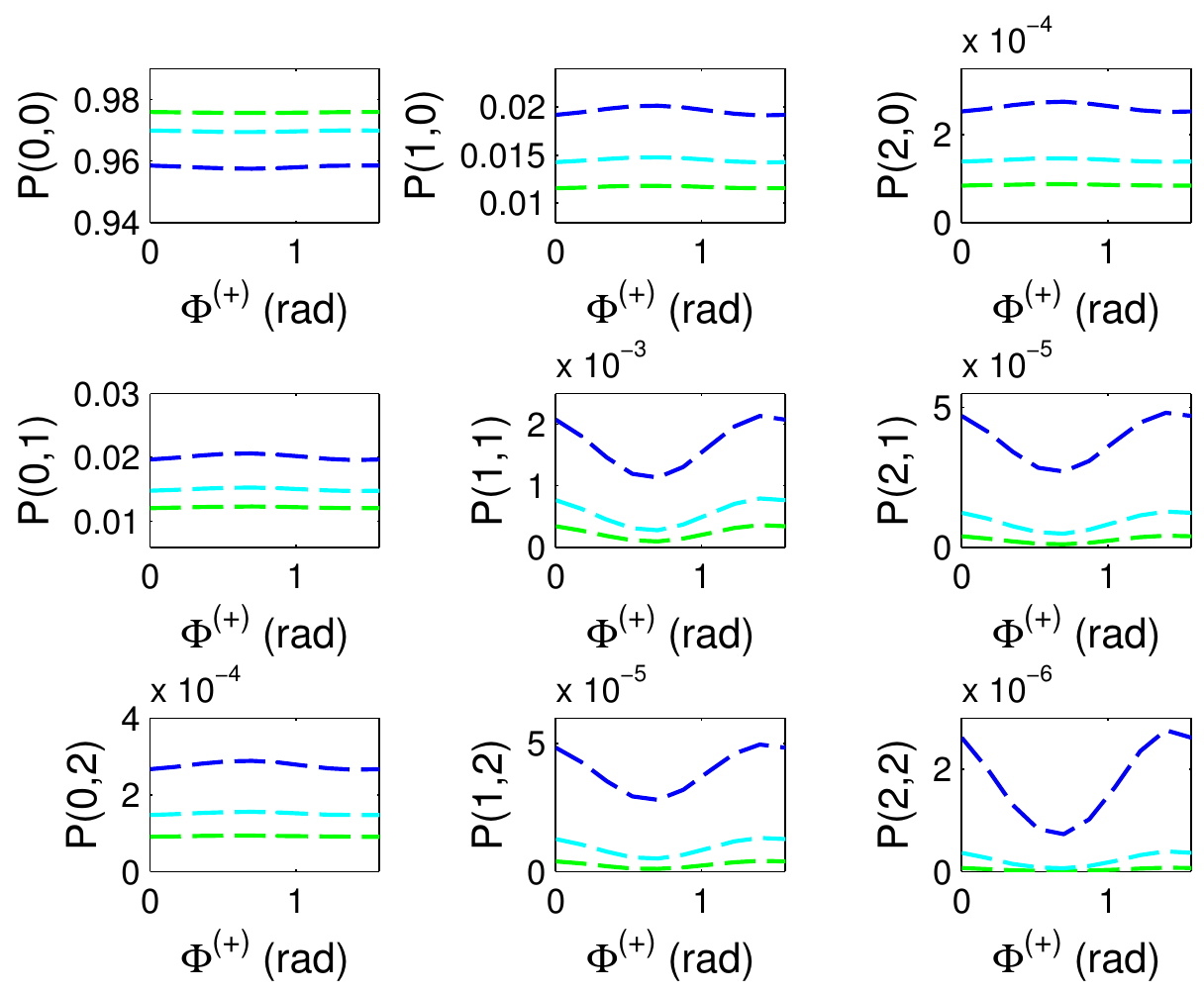}
\flushleft{{\bf Supplementary Figure 5: Theoretical joint counting statistics for different detection efficiencies in the case of a two-mode squeezed state.} We illustrate the effect of imperfect detection on the observed oscillations by varying the values of the detection efficiencies. Green curves correspond to $\eta = 0.05$, cyan to $\eta = 0.1$ and blue to $\eta = 0.2$; we assume the same efficiencies on both TMDs and use the experimentally determined values of the squeezing parameter $|\lambda|$ and weight $p$. As in the case of the SSPS, the parameter $\eta$ determines dramatic changes in the visibility of the nonclassical oscillations.}
\end{figure}


\newpage

\begin{figure}[h!!!]
\includegraphics[viewport= -100 -20 680 480, clip, width=\columnwidth]{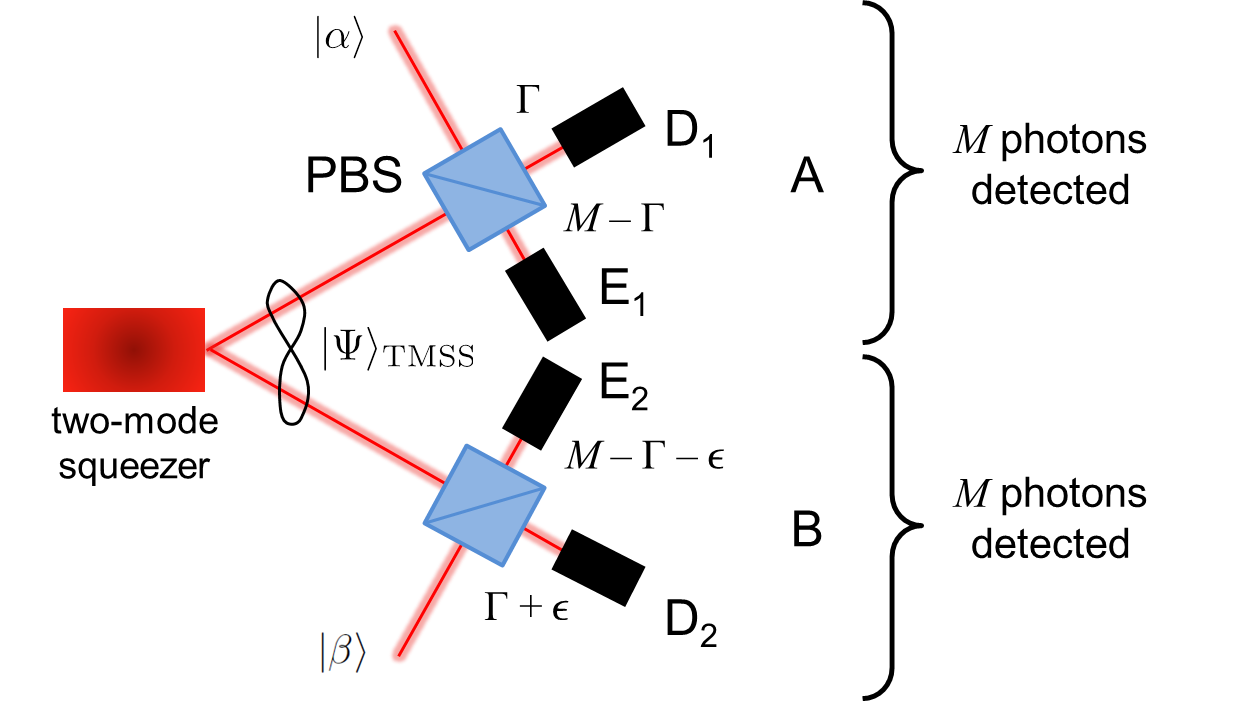}
\flushleft{{\bf Supplementary Figure 6: Study of Gaussian entanglement with weak-field homodyne (WFH) detection.} This is the ideal layout for the investigation of the nonlocal character of Gaussian entanglement via WFH. A two-mode squeezed state, an archetypal Gaussian resource, is interfered with a weak coherent state on two PBSs; photon-number-resolving detectors D$_1$, E$_1$, D$_2$, E$_2$ monitor all four outputs (\textit{i.e.}, we do not discard the reflected output modes here). If we fix the number $M$ of photons detected by each WFH detector, we can label the recorded detection events according to how $M$ particles distribute themselves on the transmitted and reflected output modes: namely, these are the quantities \, $\Gamma, M\!-\!\Gamma, \Gamma\!+\!\epsilon$ and $M\!-\!\Gamma\!-\!\epsilon$ \, shown in the diagram.}
\end{figure}

\newpage

\begin{figure}[!!!h]
\begin{center}
\includegraphics[viewport= -70 0 800 700, clip, width=\columnwidth]{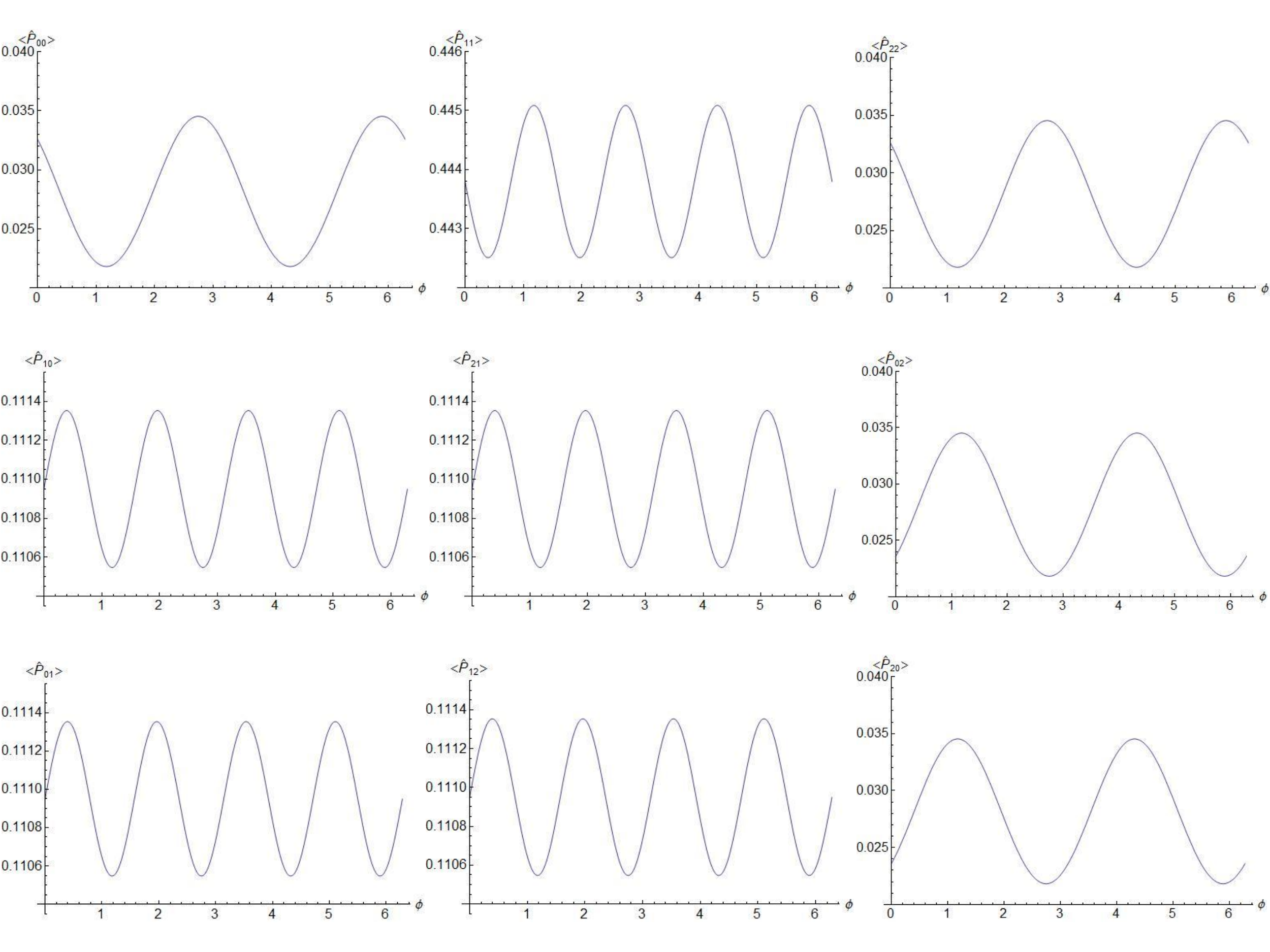}
\flushleft{{\bf Supplementary Figure 7: Probabilities of four-photon detection events as functions of the phase of the local oscillator.} The graphs show the behaviour of the expectation values of the projection operators on the considered four-photon states, \textit{i.e.}, the set $\{\expect{P_{kk^{\prime}}}\}$ for $k,k^{\prime} \leq 2$ (see Supplementary Note~3): here we take into account the multiplicative factor $\sqrt{1-|\lambda|^2}e^{-|\alpha|^2/2}e^{-|\beta|^2/2}$ neglected in the expression for $\ket{\Psi}_{\textrm{out}}$ (see Eq.\eqref{output:state}). The phase on the horizontal axis is that of either one of the local oscillators, $\phi_a$ or $\phi_b$; we set $\tilde{\theta} = 0$ for the two-mode squeezed state and $\phi_b = \phi_a + \pi/4$, relabelling $\phi_a = \phi$. Physically plausible values -- under the assumption of ideal detectors, \textit{i.e.}, $\eta = 1$ for all PNRDs -- are assigned to the other variables so that the considered functions depend solely on one phase parameter: we have $|\lambda| = 0.3$ and $|\alpha| = |\beta| = 0.131$. We also assume the PBSs to be balanced, therefore $t_i = r_j = 1/\sqrt{2}$ for $i, j = 1, 2$.}
\end{center}
\end{figure}

\newpage

\section*{Supplementary Note 1: Detection probabilities for the weak-field homodyne}
We describe the expected response of a weak-field homodyne (WFH) detector to a generic quantum state. For this purpose, we study the interference of a Fock state $\ket{n}$ with a coherent state $\ket{\alpha}$ on a beam splitter (BS) with transmittivity $t$ and reflectivity $r=\sqrt{1-t^2}$. We follow the asymmetric convention for the beam splitter input-output relations, so that the modes are transformed as $\hat{a}^{\dag}_{\textrm{in}} \rightarrow t\hat{a}^{\dag}_{\textrm{out}} + r\hat{b}^{\dag}_{\textrm{out}}$, and $\hat{b}^{\dag}_{\textrm{in}} \rightarrow t\hat{b}^{\dag}_{\textrm{out}} - r\hat{a}^{\dag}_{\textrm{out}}$. From this, it follows that the operators which generate coherent and Fock states transform as $\hat{D}_{\textrm{in}}(\alpha) \rightarrow \hat{D}_{\textrm{out}2}(t\alpha)\hat{D}_{\textrm{out}1}(-r\alpha)$ and $(\hat{a}^{\dag}_{\textrm{in}})^n \rightarrow \sum_{k=0}^{n}\binom{n}{k}t^{n-k} r^{k}(\hat{a}^{\dag}_{\textrm{out}1})^{n-k}(\hat{a}^{\dag}_{\textrm{out}2})^k$, respectively. Therefore, given the initial state
\begin{equation}
\ket{\Psi}_{\textrm{in}} = \ket{\alpha}\ket{n} = \hat{D}_{\textrm{in}}(\alpha)(\hat{a}^{\dag}_{\textrm{in}})^{n}\ket{0}\ket{0},
\end{equation}
this undergoes the above-described beam splitter transformations and becomes
\begin{equation}\label{App:tot:state}
\ket{\Psi}_{\textrm{out}} = \sum_{k=0}^{n}\binom{n}{k} t^{n-k}r^{k} \hat{D}_{\textrm{out}2}(t\alpha)\hat{D}_{\textrm{out}1}(-r\alpha)(\hat{a}^{\dag}_{\textrm{out}1})^{n-k}(\hat{a}^{\dag}_{\textrm{out}2})^{k}\ket{0}\ket{0}.
\end{equation}
We introduce the short-hand notation $\ket{\alpha;n} := \hat{D}(\alpha)(\hat{a}^{\dag})^{n}\ket{0}\ket{0}$ for the displaced Fock state; we note that the norm of $\ket{\alpha;n}$ is $n!$. For a given displacement $\alpha$, these states form an orthogonal set, $\braket{\alpha;n}{\alpha;m} = \sqrt{n!}\,\delta_{nm}$, as the displacement preserves the scalar product. This notation allows us to express the total output state in Eq.(\ref{App:tot:state}) as
\begin{equation}
\ket{\Psi}_{\textrm{out}} = \sum_{k=0}^{n} \binom{n}{k} t^{n-k}r^{k} \ket{t\alpha;n-k}\ket{{-}r\alpha;k}.
\end{equation}
As a consequence, an input state whose expression in the Fock basis is $\sum_{n=0}^{\infty}c_n \ket{n}$ transforms as 
\begin{equation}
\ket{\Psi}_{\textrm{out}} = \sum_{k=0}^{+\infty}\sum_{n=k}^{+\infty} c_{n} \binom{n}{k} t^{n-k}r^{k} \ket{t\alpha;n-k}\ket{-r\alpha;k},
\end{equation}
where we have inverted the order of the summations over $n$ and $k$. Our experimental detection scheme does not record the outcomes on mode $\hat{b}_{\textrm{out}}$, which corresponds to tracing over this mode. This leads to
\begin{equation}
\ket{\Phi}_{\textrm{out}} = \sum_{k=0}^{+\infty} r^{2k}k! \, \ket{\Phi}_{k}\,, \qquad \ket{\Phi}_{k} = \sum_{n=k}^{+\infty} c_n \binom{n}{k} t^{n-k}\ket{t\alpha;n-k}\,.
\end{equation}
The photon statistics are then given by the probability distribution
\begin{equation}\label{App:prob:distr}
\varphi(m) = \sum_{k=0}^{+\infty} r^{2k}k! \Biggl|\sum_{n=k}^{+\infty} c_{n} \binom{n}{k} t^{n-k} \braket{m}{t\alpha;n-k}\Biggr|^2\,.
\end{equation}
In order to express $\varphi(m)$ more explicitly, we recur to the relation
\begin{equation}\label{App:utility}
\hat{D}(\alpha)(\hat{a}^{\dag})^{n}\ket{0} = (\hat{a}^{\dag}{-}\alpha^{\ast})^{n}\hat{D}(\alpha)\ket{0}\,,
\end{equation}
which is verified by virtue of the unitarity of the displacement operator $\hat{D}$, along with the relation $\hat{D}^{\dag}(\alpha)\hat{a}^{\dag}\hat{D}(\alpha) = \hat{a}^{\dag} + \alpha^{\ast}$. \\
Consequently, the inner product $\braket{m}{\alpha;n}$ which appears in Eq.\eqref{App:prob:distr} can be written as
\begin{equation}
\begin{split}
\braket{m}{\alpha;n} &= \bra{m}\sum_{\kappa=0}^{n} \binom{n}{\kappa} (\hat{a}^{\dag})^{\kappa}(-\alpha^{\ast})^{n-\kappa} \ket{\alpha} = \\ &= \sum_{\kappa=0}^{n} \binom{n}{\kappa} \frac{\sqrt{m!}}{\sqrt{(m-\kappa)!}}(-1)^{n-\kappa}|\alpha|^{m+n-2\kappa}e^{i(m-n)\phi}e^{-|\alpha|^2/2},
\end{split}
\end{equation}
where $\alpha = |\alpha|e^{i\phi}$. \\
Finally, we have to take into account the response of the time-multiplexed detector (TMD). This is characterised by its non-unit detection efficiency $\eta$, which we model as a beam splitter with reflectivity $\eta$, and draws its photon-number resolution from temporal binning. These two aspects -- loss and binned photon counting -- are accounted for by two matrices: $L(\eta)$ models photon loss, while $C$ describes the binning operation~\cite{Achilles1,Zhang,Worsley}. If we denote by ${\bf f} = \{\varphi(0),\varphi(1),\varphi(2),...\}$ the vector of photon statistics and by ${\bf p} = \{p(0),p(1),p(2),...\}$ the vector of counting statistics, the following relation holds:
\begin{equation}
{\bf p} = C\cdot L(\eta) \cdot {\bf f}.
\end{equation}
This single-mode expression can be easily generalised to our two-mode layout. The joint photon statistics for two WFH detectors with settings $\{r,t,\alpha\}$ and $\{r',t',\alpha'\}$, observing the state $\sum_{n=0}^{+\infty}c_{n,n'} \ket{n}\ket{n'}$, are given by
\begin{equation}
\Phi(m,m^{\prime}) = \sum_{k=0}^{+\infty}\sum_{k^{\prime}=0}^{+\infty} k!k^{\prime}!\,r^{2k}(r^{\prime})^{2k^{\prime}} \Biggl|\sum_{n=k}^{+\infty}\sum_{n^{\prime}=k^{\prime}}^{+\infty} c_{nn^{\prime}} \binom{n}{k}\binom{n^{\prime}}{k^{\prime}} t^{n-k}(t^{\prime})^{n^{\prime}-k^{\prime}} \braket{m}{t\alpha;n-k}\braket{m^{\prime}}{t^{\prime}\alpha^{\prime};n^{\prime}-k^{\prime}}\Biggr|^2\,.
\end{equation}
Loss on each mode -- described by the efficiencies $\eta_1$ and $\eta_2$, respectively -- and temporal binning can be taken into account by casting the probabilities $\Phi(m,m')$ into a matrix $\underline{\underline F}$, which is related to the matrix of the joint detection probabilities $\underline{\underline{P}}$ through~\cite{Worsley}
\begin{equation}
\label{theory:P}
\underline{\underline{P}} = C_1 \cdot L(\eta_1) \cdot\underline{\underline{F}} \cdot L(\eta_2)^{{\rm T}} \cdot C_{2}^{{\rm T}}\,.
\end{equation}

\section*{Supplementary Note 2: Theoretical model of the experiment}
We adopt a collinear geometry for our WFH detectors in order to simplify the experimental scheme and ensure better passive phase stability (see Supplementary Fig.~1 and Supplementary Fig.~3). The relative phase between signal and local oscillator is varied by means of birefringent elements arranged in a geometric phase rotator (GPR). However, imperfections in the individual wave plates and calibration inaccuracies can determine an undesired modulation of the reflectivity associated to the polarising beam splitter in each WFH detector. We thus perform a calibration of both GPRs in the presence of the local oscillator only. A typical plot of the dependence of counts on the setting of the GPR is shown in Supplementary Fig.~2. We account for this spurious effect by expressing the reflectivity as $r(\Theta)=r_0\sqrt{1+v \cos(4\Theta+\theta_0)}$, where $r_0 = 0.5$ is the ideal reflectivity, $\Theta$ is the setting of the half-wave plate in the GPR, $\theta_0$ is an offset angle, and the modulation depth $v$ is expected to be of the order of $0.1$ for small deviations from the perfect case. This relation for $r(\Theta)$ is included in Eq.\eqref{theory:P} for the expected joint detection probabilities.

\medskip

\noindent{\it Modelling a split single-photon state (SSPS).} We want to compare our experimentally generated split single photon to a theoretical signal. It is reasonable to assume that the SSPS exhibits some degree of mixedness~\cite{Bartley1}, therefore we take into account imperfect single-photon generation by including contributions from the vacuum and higher-order terms from the downconversion process (given the brightness of our source) in the expression of our state:
\begin{equation}\label{SI:rhoSSPS}
\rho_0=w_0\ket{0}\bra{0}+w_1\ket{1}\bra{1}+(1-w_0-w_1)\ket{2}\bra{2}.
\end{equation}
This state produces an entangled resource when it is split on a symmetric beam splitter. The corresponding counting statistics read
\begin{equation}\label{SI:SSPSCntsMatrix}
\underline{\underline{P}}=w_0\underline{\underline{P_0}}+w_1\underline{\underline{P_1}}+(1-w_0-w_1)\underline{\underline{P_2}},
\end{equation}
where the matrix $\underline{\underline{P_k}}$ describes the statistics associated with the entangled state obtained from Fock layer $\ket{k}$.

The coefficients $w_0$ and $w_1$, along with the detection efficiencies $\eta_{\rm A}$ and $\eta_{\rm B}$, are the key parameters which appear in our model. The latter are determined from the counting statistics of the experimental input state when no local oscillator impinges on the PBS, following the Klyshko calibration method. In order to compute $w_0$ and $w_1$, we suitably invert Eq.(\ref{theory:P}) and obtain the photon statistics matrix of the source, which we label $\underline{\underline{F}}^{\textrm{PDC}}$; we thus estimate $w_0 = \underline{\underline{F}}^{\textrm{PDC}}(0,0)$ and $w_1 = \underline{\underline{F}}^{\textrm{PDC}}(1,0) + \underline{\underline{F}}^{\textrm{PDC}}(0,1)$.

In Supplementary Fig.~4 we show the theoretical curves for the joint counting statistics matrix $\underline{\underline{P}}$ up to two detection events on each output mode. These plots inform us about the main features of the physical phenomenon which we probe with our WFH detectors: coherent oscillations across Fock layers are present in all terms but those where at least one side detects no photons. Moreover, the oscillations are shown to be in phase with one another. As a further investigation of the role of loss in our scheme (which we have already taken into account by using the experimental detection efficiencies in our model), Supplementary Fig.~4 includes three distinct detection scenarios given by different values of $\eta_{\rm A}$ and $\eta_{\rm B}$; poor efficiency of detection appears to mainly affect the visibility of the studied nonclassical oscillations.

\medskip

\noindent{\it Modelling a two-mode squeezed state (TMSS).}
 The expression for an ideal two-mode squeezed vacuum state reads
\begin{equation}\label{SI:TMSS}
\ket{\Psi}_\textrm{TMSS}=\sqrt{1-|\lambda|^2}\sum_{n=0}^{+\infty}\lambda^n\ket{n}_\textrm{A}\ket{n}_\textrm{B},
\end{equation}
where $|\lambda|$ is the real squeezing parameter determined by the brightness of the parametric downconversion source. However, a more realistic expression for the experimentally produced state is 
\begin{equation}\label{SI:rhoTMSS}
\rho = (1-p)\ket{\Psi}_\textrm{TMSS}\bra{\Psi}+p\ket{0,1}\bra{0,1},
\end{equation}
where the term $\ket{0,1}\bra{0,1}$ is a first-order approximation of noise in a squeezed thermal state. Indeed, the presence of such contribution is suggested by the experimental $g^{(2)}$ for the source, whose values highlight the two following features for our squeezer: an asymmetry across the two modes, and the presence of unwanted frequency modes in the generated two-mode squeezed state. Both effects are captured by the additional term appearing in Eq.(\ref{SI:rhoTMSS}).

In this case, the relevant quantities for the theoretical model are given by the squeezing parameter and the weight $p$. The former can be computed from the photon statistics of the downconversion source; manual inspection of the experimental and theoretical plots suggests the value $p = 0.04$ for the coefficient in the expression of $\rho$.

The theoretical curves for the joint counting statistics matrix $\underline{\underline{P}}$ (up to two detection events) for the TMSS are presented in Supplementary Fig.~5. Similarly to the case of a split single photon, the model predicts coherent, in-phase oscillations across the Fock layers for all terms but those where at least one side detects vacuum. Once again, the effect of loss is a reduction of the visibility of such oscillations.

\bigskip

As already pointed out in the main text, we remark that our theoretical models the SSPS and the TMSS share some common limitations. Notably, imperfect mode matching affects the interference between the probed state and the classical reference. For this reason, this aspect is likely to act adversely on the joint detection statistics too. Nevertheless, our descriptions have the advantage of relying on very few parameters, most of which are directly estimated from the collected data. Each model agrees with the data for all considered terms in the joint counting statistics simultaneously, which in turn means that we can harness the behaviour of all multi-photon components in a given output state with one single description and set of parameters.

\section*{Supplementary Note 3: Nonlocality of a two-mode squeezed state detected with weak-field homodyne}
In this section we provide details on the application of the Collins-Gisin-Linden-Massar-Popescu (CGLMP) inequalities~\cite{Collins} to WFH. Following the notation introduced in Supplementary Fig.~6, we illustrate the case in which the number of detected photons on each side is $M = 2$, {\it i.e.}, we restrict our study to local subsystems with dimension $D = M+1 = 3$. The generalisation to higher photon numbers proceeds in a similar fashion. Let us consider the two-mode quantum state $\ket{\Psi}_{\rm TMSS}$, distributed across modes $\hat{a}^\dag_{\rm s},\hat{a}^\dag_{\rm i}$, and the two local oscillators $\ket{\alpha}$ and $\ket{\beta}$ on modes $\hat{b}^\dag_1$ and $\hat{b}^\dag_2$ respectively. As we focus on the detection of low photon numbers ($M = 2$), we write the input state as
\begin{equation}
\begin{split}
\ket{\Psi}_{\rm in} = \Biggl(\frac{\alpha^2}{2}\frac{\beta^2}{2}(\hat{b}^{\dag}_{1})^{2}(\hat{b}^{\dag}_{2})^{2} + \frac{\lambda^2}{2}(\hat{a}^{\dag}_{\rm s})^{2}(\hat{a}^{\dag}_{\rm i})^{2} + \lambda\alpha\beta\hat{a}^{\dag}_{\rm s}\hat{a}^{\dag}_{\rm i}\hat{b}^{\dag}_{1}\hat{b}^{\dag}_{2} \Biggr)\ket{0}_{\rm s}\ket{0}_{\rm i}\ket{0}_{1}\ket{0}_{2},
\end{split}
\end{equation}
where $\alpha = |\alpha|e^{\imath \phi_a}$, $\beta = |\beta|e^{\imath \phi_b}$ and $\lambda = |\lambda|e^{\imath \tilde{\theta}}$. \\
The BS of each WFH detector operates the following transformation from the input modes into output modes $\{\hat{d}_{1}^{\dag},\hat{e}_{1}^{\dag},\hat{d}_{2}^{\dag},\hat{e}_{2}^{\dag}\}$:
\begin{equation}\label{I:O:BS}
\left\{\begin{array}{l} \hat{a}_{\rm s}^{\dag} = t_{1} \hat{d}_{1}^{\dag} + r_{1}^{\ast} \hat{e}_{1}^{\dag}\,, \\ \hat{b}_{1}^{\dag} = r_{1} \hat{d}_{1}^{\dag} - t_{1} \hat{e}_{1}^{\dag}\,, \\ \\ \hat{a}_{\rm i}^{\dag} = t_{2} \hat{d}_{2}^{\dag} + r_{2}^{\ast} \hat{e}_{2}^{\dag}\,, \\ \hat{b}_{2}^{\dag} = r_{2} \hat{d}_{2}^{\dag} - t_{2} \hat{e}_{2}^{\dag}\;,
\end{array} \right.
\end{equation}
where we assume that $t_{j} \in \mathbb{R}$ and $r_{j} = -\imath|r_{j}|$ for $j = 1, 2$. \\
These relations lead to the expression for the output state $\ket{\Psi}_{\rm out}$ in the Fock basis (up to a normalisation factor):
\begin{equation}\label{output:state}
\begin{split}
\ket{\Psi}_{\rm out} &= \Bigl(\frac{\alpha^2 \beta^2}{4}r_{1}^{2}r_{2}^{2} + \frac{\lambda^2}{2}t_{1}^{2}t_{2}^{2} + \lambda\alpha\beta t_{1}t_{2}r_{1}r_{2}\Bigr)\ket{2200} + \Bigl(\frac{\alpha^2 \beta^2}{4}t_{1}^{2}t_{2}^{2} + \frac{\lambda^2}{2}(r_{1}^{\ast})^{2}(r_{2}^{\ast})^{2} + \lambda\alpha\beta t_{1}t_{2}r_{1}^{\ast}r_{2}^{\ast}\Bigr)\ket{0022} \; + \\ &\quad  + \Bigl(\frac{\alpha^2 \beta^2}{4}r_{1}^{2}t_{2}^{2} + \frac{\lambda^2}{2}t_{1}^{2}(r_{2}^{\ast})^{2} - \lambda\alpha\beta t_{1}t_{2}r_{1}r_{2}^{\ast}\Bigr)\ket{2002} + \Bigl(\frac{\alpha^2 \beta^2}{4}t_{1}^{2}r_{2}^{2} + \frac{\lambda^2}{2}(r_{1}^{\ast})^{2}t_{2}^{2} - \lambda\alpha\beta t_{1}t_{2}r_{1}^{\ast}r_{2}\Bigr)\ket{0220} \; + 
\\ &\quad + \bigl[\alpha^2 \beta^2 t_{1}t_{2}r_{1}r_{2} + 2\lambda^2 t_{1}t_{2}r_{1}^{\ast}r_{2}^{\ast} - \lambda\alpha\beta(t_{2}^{2}|r_1|^2 - |r_1|^2|r_2|^2 + t_{1}^{2}|r_2|^2 - t_{1}^{2}t_{2}^{2})\bigr]\ket{1111} \; + \\ &\quad + \Bigl[-\frac{\alpha^2 \beta^2}{2}t_{2}r_{1}^{2}r_{2} + \lambda^{2}t_{1}^{2}t_{2}r_{2}^{\ast} - \lambda\alpha\beta (t_{1}t_{2}^{2}r_{1} - t_{1}r_{1}|r_2|^2)\Bigr]\ket{2101} \; + \\ &\quad + \Bigl[-\frac{\alpha^2 \beta^2}{2}t_{1}r_{1}r_{2}^{2} + \lambda^{2}t_{1}t_{2}^{2}r_{1}^{\ast} - \lambda\alpha\beta(t_{1}^{2}t_{2}r_{2} - t_{2}|r_1|^{2}r_{2})\Bigr]\ket{1210} \; + \\ &\quad + \Bigl[-\frac{\alpha^2 \beta^2}{2}t_{1}^{2}t_{2}r_{2} + \lambda^{2}t_{2}(r_{1}^{\ast})^{2}r_{2}^{\ast} - \lambda\alpha\beta(t_{1}r_{1}^{\ast}|r_2|^2 - t_{1}t_{2}^{2}r_{1}^{\ast})\Bigr]\ket{0121} \; + \\ &\quad + \Bigl[-\frac{\alpha^2 \beta^2}{2}t_{1}t_{2}^{2}r_{1} + \lambda^{2}t_{1}r_{1}^{\ast}(r_{2}^{\ast})^{2} + \lambda\alpha\beta(t_{1}^{2}t_{2}r_{2}^{\ast} - t_{2}|r_1|^{2}r_{2}^{\ast})\Bigr]\ket{1012}.
\end{split}
\end{equation}
Here we use the convention $\ket{0000} = \ket{0}_{1{\rm t}}\ket{0}_{2{\rm t}}\ket{0}_{1{\rm r}}\ket{0}_{2{\rm r}}$, and similarly for the other number states (`t' and `r' denoting transmitted and reflected output modes, respectively).

In the work of Collins and collaborators, the local realistic bound is calculated on a linear combination of detection probabilities associated to local measurement settings -- in our case these are the complex parameters $\alpha$ and $\beta$. As introduced in~\cite{Collins}, we adopt the shorthand notation $P(A_i = B_j + \epsilon)$ to indicate the probability of an event for which the detection outcomes $A_i$ and $B_j$ differ by $\epsilon = -1, 0, 1$ (where $\{-1,0,1\}$ and $\{0,1,2\}$ are congruent modulo $3$). Therefore, the CGLMP inequality adapted to our layout reads
\begin{equation*}
|I_3| \leq 2\:,
\end{equation*}
where
\begin{equation}\label{I3}
\begin{split}
I_3 = &[P(A_1 = B_1) + P(B_1 = A_2 + 1) + P(A_2 = B_2) + P(B_2 = A_1)] - \\ &- [P(A_1 = B_1 - 1) + P(B_1 = A_2) + P(A_2 = B_2 - 1) + P(B_2 = A_1 - 1)].
\end{split}
\end{equation}
According to the notation introduced above, each term in Eq.\eqref{I3} is computed as follows:
\begin{equation}\label{probs:outcomes}
\begin{split}
&P(A_1 = B_1) = \expect{\hat{P}_{00}}(\alpha_{1},\beta_{1}) + \expect{\hat{P}_{11}}(\alpha_{1},\beta_{1}) + \expect{\hat{P}_{22}}(\alpha_{1},\beta_{1}), \\
&P(B_1 = A_2 + 1) = \expect{\hat{P}_{10}}(\alpha_{2},\beta_{1}) + \expect{\hat{P}_{21}}(\alpha_{2},\beta_{1}) + \expect{\hat{P}_{02}}(\alpha_{2},\beta_{1}), \\
&P(A_2 = B_2) = \expect{\hat{P}_{00}}(\alpha_{2},\beta_{2}) + \expect{\hat{P}_{11}}(\alpha_{2},\beta_{2}) + \expect{\hat{P}_{22}}(\alpha_{2},\beta_{2}), \\
&P(B_2 = A_1) = \expect{\hat{P}_{00}}(\alpha_{1},\beta_{2}) + \expect{\hat{P}_{11}}(\alpha_{1},\beta_{2}) + \expect{\hat{P}_{22}}(\alpha_{1},\beta_{2}), \\
&P(A_1 = B_1 - 1) = \expect{\hat{P}_{10}}(\alpha_{1},\beta_{1}) + \expect{\hat{P}_{21}}(\alpha_{1},\beta_{1}) + \expect{\hat{P}_{02}}(\alpha_{1},\beta_{1}), \\
&P(B_1 = A_2) = \expect{\hat{P}_{00}}(\alpha_{2},\beta_{1}) + \expect{\hat{P}_{11}}(\alpha_{2},\beta_{1}) + \expect{\hat{P}_{22}}(\alpha_{2},\beta_{1}), \\
&P(A_2 = B_2 - 1) = \expect{\hat{P}_{10}}(\alpha_{2},\beta_{2}) + \expect{\hat{P}_{21}}(\alpha_{2},\beta_{2}) + \expect{\hat{P}_{02}}(\alpha_{2},\beta_{2}), \\
&P(B_2 = A_1 - 1) = \expect{\hat{P}_{01}}(\alpha_{1},\beta_{2}) + \expect{\hat{P}_{12}}(\alpha_{1},\beta_{2}) + \expect{\hat{P}_{20}}(\alpha_{1},\beta_{2}),
\end{split}
\end{equation}
where we conveniently relabelled the four-photon output states. Namely,
\begin{equation*}
\begin{split}
&\ket{2200} \longleftrightarrow \ket{00} \rightarrow \hat{P}_{00}\:, \quad \ket{1111} \longleftrightarrow \ket{11} \rightarrow \hat{P}_{11}\:, \\
&\ket{0022} \longleftrightarrow \ket{22} \rightarrow \hat{P}_{22}\:, \quad \ket{1210} \longleftrightarrow \ket{10} \rightarrow \hat{P}_{10}\:, \\
&\ket{0121} \longleftrightarrow \ket{21} \rightarrow \hat{P}_{21}\:, \quad \ket{2002} \longleftrightarrow \ket{02} \rightarrow \hat{P}_{02}\:, \\
&\ket{2101} \longleftrightarrow \ket{01} \rightarrow \hat{P}_{01}\:, \quad \ket{1012} \longleftrightarrow \ket{12} \rightarrow \hat{P}_{12}\:, \\
&\ket{0220} \longleftrightarrow \ket{20} \rightarrow \hat{P}_{20}\:.
\end{split}
\end{equation*}
This operation amounts to retaining the number of detected photons on the two reflected output modes as the two significant digits for the labels. The expectation values $\expect{\cdot}$ of the projection operators $\{\hat{P}_{kk^{\prime}}\}$, $k, k^{\prime} = 0, 1, 2$, appear in Eq.(\ref{probs:outcomes}). These are calculated on the output state $\ket{\Psi}_{\rm out}$, {\it i.e.}, $\expect{\hat{P}_{kk^{\prime}}} = \bra{\Psi}\hat{P}_{kk^{\prime}}\ket{\Psi}_{\rm out}$ for all $k, k^{\prime}$ considered.  The quantities $\{\alpha_{l},\beta_{l^{\prime}}\}$, $l, l^{\prime} = 1, 2$, refer to physical measurement settings, which in our case are identified by the complex amplitude of each local oscillator.

In Supplementary Fig.~7 we show the oscillations of the terms $\{\expect{\hat{P}_{kk^{\prime}}}\}$, $k, k^{\prime} = 0, 1, 2$, which appear in Eq.\eqref{I3} through the detection probabilities $P(A_i = B_j + \epsilon)$, as the phase of one local oscillator is varied. Once again, we note that this study assumes an ideal setup, from the generation of pure two-mode squeezed states to perfect WFH detectors. The most remarkable feature in Supplementary Fig.~7 is the presence of two different oscillation periods for $\{\expect{\hat{P}_{kk^{\prime}}}\}$, the period being determined by the specific Fock layer observed.

The presence of different oscillation periods associated to distinct photon numbers has been observed when a single-mode squeezed state is interfered with a coherent state of comparable energy~\cite{Afek}. However, the presence of such super-resolved fringes does not imply a quantum enhancement in phase estimation. In fact, while super-resolution can be obtained by purely classical means~\cite{Resch,Kothe}, super-sensitivity can only be achieved with quantum resources. An analysis in this direction needs to account for all resources and all post-selection probabilities. The usefulness of the scheme discussed in this work for quantum metrology is currently under investigation.

Finally, we comment on the role of detection efficiency in the framework of the generalised Bell inequalities we have just presented. As shown in Supplementary Fig.~4 and Supplementary Fig.~5, lower detection efficiencies decrease the visibility of the oscillations and artificially enhance the low-photon components. The effect is clearly more detrimental than the induction of a detection loophole, as in standard Bell tests performed with entangled photons. \\
Further, this effect is more pronounced for higher Fock layers. In fact, the first-order contribution to noise in Fock layer $M$ comes from the higher layer $M+1$. Each term in Eq.(\ref{I3}) may then be written as
\begin{equation}
P\propto \eta^M P_M + (M{+}1) \eta^M(1-\eta) \nu_{M{+}1}
\end{equation}
where $P_M$ is the correlation in the ideal case ${\eta}{=}1$, and $\nu_{M+1}$ is a noise term coming from Fock layer $M+1$. The latter is due to the loss of a single photon, which one could otherwise reject by post-selection. We thus see that the relative weight becomes more important as $M$ increases. Therefore, the achievable violation not only decreases with $M$, but its resilience to loss also becomes weaker.

\end{document}